\documentclass[final,3p,twocolumn]{elsarticle}

\usepackage[hyphens]{url}

\usepackage{lineno}
\usepackage[hidelinks]{hyperref}

\usepackage{cancel}
\usepackage{pgfplotstable}
\usepackage{tabularx}
\usepackage{caption}
\usepackage{amsmath}
\usepackage{comment}
\usepackage[export]{adjustbox}
\usepackage[official]{eurosym}

\usepackage{ctable}
\usepackage[nonumberlist]{glossaries}
\usepackage{siunitx}
\usepackage{array}
\usepackage{afterpage}

\usepackage{listings}
\usepackage{graphicx}
\usepackage{caption}
\usepackage{subcaption}

\makeglossaries
\usepackage[capitalise]{cleveref}

\modulolinenumbers[5]

\journal{Solar Energy}

\glsdisablehyper

\newglossaryentry{gls:tau_time}{name={$\tau$},description={Time interval [second, hour, day, year]}
}
\newglossaryentry{gls:PR_tau}{name={$PR_{\tau}$},description={Performance Ratio over period $\tau$}
}
\newglossaryentry{gls:Final_Yield}{name={ $Y_{f,\tau}$ },
description={%
Final yield over period $\tau$ [\si[per-mode=symbol]{kWh \per kW}$_p\cdot \tau$]} 
}

\newglossaryentry{gls:Reference_Yield}{name={$Y_{r,\tau}$},
description={%
Reference yield over period $\tau$ [\si[per-mode=symbol]{kWh \per m^2 }$\cdot \tau$]}
} 
\newglossaryentry{gls:G}{name={$G$},description={Irradiance [\si[per-mode=symbol]{W \per m^2}]}
}
\newglossaryentry{gls:G_STC}{name={$G_{STC}$},
description={STC Irradiance: 1000 [\si[per-mode=symbol]{W \per m^2}]}
}
\newglossaryentry{gls:STC}{name={STC},description={Standardized Testing Conditions}
}
\newglossaryentry{gls:Eta_Norm_non_TC_tau}{name={$\langle\eta_{N}\rangle_{\tau}$},
description={ %
Normalized average efficiency of PV module or system over period $\tau$ [-]}
}
\newglossaryentry{gls:Eta_Norm_TC_tau}{name={$\langle\eta_{N}^*\rangle_{\tau}$},
description={ %
Normalized average temperature-corrected efficiency of PV module or system over period $\tau$ [-]}
}
\newglossaryentry{gls:Eta_Norm_func_time_non_TC}{name={$\eta_{N}(t)$},description={ %
Normalized instantaneous PV module or array efficiency [-]}
}
\newglossaryentry{gls:Eta_Norm_func_time_TC}{name={$\eta_{N}^*(t)$},description={ %
Normalized temperature-corrected instantaneous PV module or array efficiency [-]}
}
\newglossaryentry{gls:Power_STC}{name={$P_{STC}$},description={STC Power of PV module or array [\si{kW_p}]}
}
\newglossaryentry{gls:gamma_power}{name={$\gamma$},description={Efficiency temperature coefficient [\si[per-mode=symbol]{\% \per \degreeCelsius}]}
}
\newglossaryentry{gls:T_STC}{name={$T_{STC}$},description={STC module cell temperature: 25 [\si{\degreeCelsius}]}
}
\newglossaryentry{gls:Delta_T_STC}{name={$\Delta T_{STC}$},description={ %
Module temperature difference to $T_{STC}$ [\si{\degreeCelsius}]}
}
\newglossaryentry{gls:P_norm}{name={$P_N$},description={Normalized power of module or system [\si{kW$_p$}]}
}
\newglossaryentry{gls:Delta_P_norm}{name={$\Delta P_N$},description={ %
Normalized loss from irradiance to normalized power [\si[per-mode=symbol]{W_p}]}
}

\newglossaryentry{gls:G_G_STC_frac}{name={$G'$},description={Fraction of $G/G_{STC}$ [-]}
}
\newglossaryentry{gls:MPP}{name={MPP},description={Maximum Power Point}
}
\newglossaryentry{gls:Delta_G_standard_dev}{name={$\sigma_{\Delta G}$},description={Standard deviation of the irradiance rate of change [\si[per-mode=symbol]{W \per m^2}]}
}
\newglossaryentry{gls:P_Norm_PV}{name={$P_N$},description={Normalized PV module or system power [\si{kW_p}]}
}
\newglossaryentry{gls:NVI}{name={NVI},description={Natural Variability Index [-]}
}
\newglossaryentry{gls:P_modelled_temp_corr}{name={$P_{N,mod}^*$},description={Normalized temperature-corrected modelled power}}

\newacronym{glsac:NVI}{NVI}{Natural Variability Index}
\newacronym{glsac:P_modelled_temp_corr}{$P_{N,mod}^*$}{normalized temperature-corrected modelled power}
\newacronym{glsac:SumSquareError}{SSE}{Square root of the Sum of Errors squared}

\newglossaryentry{gls:tau}{name={$\tau$},description={Thermal time constant [s]}
}

\newglossaryentry{gls:tau_electrical}{name={$\tau_e$},description={Electrical time constant [s]}
}

\newglossaryentry{gls:Tover}{name={$T_o$},description={Over-temperature versus ambient, i.e. $\Delta T_{module-ambient}$ [\si{K}]}
}

\newglossaryentry{gls:T_cell}{name={$T_{cell}$},description={Cell temperature [\si{\degreeCelsius}]}
}

\newglossaryentry{gls:T_module}{name={$T_{m}$},description={Module temperature (cell or backsheet) [\si{\degreeCelsius}]}
}

\newglossaryentry{gls:T_BS}{name={$T_{BS}$},description={Backsheet temperature [\si{\degreeCelsius}]}
}

\newglossaryentry{gls:R_wind_equiv}{name=$R_{W}$, description={Wind-affected thermal equivalent resistance [\si{K\per W}] }}

\newglossaryentry{gls:r_eq_per_m2}{name=$r_{eq}$, description={Total equivalent resistance per unit area, or R-value [\si{K\per (W/m^2)}] } }
\newglossaryentry{gls:r_eq_module_per_m2}{name=$r_{M}$, description={Module equivalent resistance per unit area [\si{K\per (W/m^2)}] } }
\newglossaryentry{gls:r_film_per_m2}{name=$r_{film}$, description={Air film equivalent resistance per unit area [\si{K\per (W/m^2)}] } }

\newglossaryentry{gls:c_eq_per_m2}{name=$c_{eq}$, description={Total equivalent capacitance per unit area, or C-value [\si{J\per (K\cdot m^2)}] } }
\newglossaryentry{gls:c_eq_module_per_m2}{name=$c_{M}$, description={Module equivalent capacitance per unit area [\si{J\per (K\cdot m^2)}] } }
\newglossaryentry{gls:c_film_per_m2}{name=$c_{film}$, description={Air film equivalent capacitance per unit area [\si{J\per (K\cdot m^2)}] } }

\newglossaryentry{gls:Ta}{name=$T_{a}$, description={Ambient temperature [\si{\degreeCelsius}]} }
\newglossaryentry{gls:Tsky}{name=$T_{sky}$, description={Sky temperature [\si{K}]} }

\newglossaryentry{gls:L_thickness}{name=$L$, description={Material thickness [\si{m}]} }
\newglossaryentry{gls:lambda_conductivity}{name=$\lambda$, description={Material thermal conductivity [\si{W/(m\cdot K)}]} }
\newglossaryentry{gls:Area}{name=$A$, description={Surface area [\si{m^2}]} }
\newglossaryentry{gls:rho_density}{name=$\rho$, description={Material density [\si{kg/m^3}] } }
\newglossaryentry{gls:specific_heat_capacity}{name=$c_p$, description={Material specific heat capacity [\si{J/(kg\cdot K)}] } }

\newglossaryentry{gls:k}{name=$k$, description={Ross coefficient [\si{K\per (W/m^2)}] } }
\newglossaryentry{gls:k_Ross}{name=$k_{Ross}$, description={Ross coefficient for all wind speeds [\si{K\per (W/m^2)}] } }

\newglossaryentry{gls:k_BS_C}{name=$k_{BS-C}$, description={Ross coefficient for backsheet-to-cell temperature correction [\si{K\per (W/m^2)}] } }

\newglossaryentry{gls:k_W}{name=$k_{W}$, description={Irradiance convection coefficient, used for WM2 [\si{K \cdot m \cdot s\per W}] } }

\glsadd{gls:tau}
\glsadd{gls:tau_electrical}
\glsadd{gls:Ta}
\glsadd{gls:T_cell}
\glsadd{gls:T_BS}
\glsadd{gls:T_module}
\glsadd{gls:Tsky}
\glsadd{gls:Tover}

\glsadd{gls:r_eq_per_m2}
\glsadd{gls:r_eq_module_per_m2}
\glsadd{gls:r_film_per_m2}

\glsadd{gls:c_eq_per_m2}
\glsadd{gls:c_eq_module_per_m2}
\glsadd{gls:c_film_per_m2}

\glsadd{gls:L_thickness}
\glsadd{gls:lambda_conductivity}
\glsadd{gls:Area}
\glsadd{gls:rho_density}
\glsadd{gls:specific_heat_capacity}

\glsadd{gls:k}
\glsadd{gls:k_W}

\begin{document}

\begin{frontmatter}

\title{Investigating methods to improve photovoltaic thermal models at second-to-minute timescales} 

\author[label1]{Bert Herteleer\corref{cor1}}
\ead{bert.herteleer@kuleuven.be}
\author[label1]{Anastasios Kladas}
\ead{anastasios.kladas@kuleuven.be}
\author[label2,label3]{Gofran Chowdhury}
\ead{gofran.chowdhury@kuleuven.be}
\author[label3,label2]{Francky Catthoor}
\ead{francky.catthoor@imec.be}
\author[label1]{Jan Cappelle}
\ead{jan.cappelle@kuleuven.be}

\address[label1]{ELECTA Gent, Faculty of Technology Engineering, KU Leuven, Gebroeders De Smetstraat 1, 9000 Gent, Belgium}
\cortext[cor1]{Corresponding author}

\address[label2]{KU Leuven, ELECTA/ ESAT – MICAS, Kasteelpark Arenberg 10, 3001 Leuven, Belgium}
\address[label3]{Imec, Kapeldreef 75, 3001 Heverlee, Belgium}

\begin{abstract}
This paper presents a range of methods to improve the accuracy of equation-based thermal models of PV modules at second-to-minute timescales. We present an RC-equivalent conceptual model for PV modules, where wind effects are captured. We show how the thermal time constant $\tau$ of PV modules can be determined from measured data, and subsequently used to make static thermal models dynamic by applying the Exponential Weighted Mean (EWM) approach to irradiance and wind signals. On average, $\tau$ is \SI{6.3 \pm 1}{min} for fixed-mount PV systems. Based on this conceptual model, the Filter- EWM - Mean Bias Error correction (FEM) methodology is developed. We propose two thermal models, WM1 and WM2, and compare these against the models of Ross, Sandia, and Faiman on twenty-four datasets of fifteen sites, with time resolutions ranging from \SI{1}{s} to \SI{1}{h}, the majority of these at \SI{1}{min} resolution. The FEM methodology is shown to reduce model errors (RMSE and MAE) on average for all sites and models versus the standard steady-state equivalent by \SI{-1.1}{K} and \SI{-0.75}{K} respectively. 
\end{abstract}

\begin{keyword}
\texttt{dynamic thermal model\sep time constant \sep KPIs}
\end{keyword}

\end{frontmatter}


\section{Introduction}

In light of the continued and increasing deployment of PV systems worldwide and their increasing importance to power grids \cite{Herteleer_Visions_future_curtailment_flexibility_PVSEC_2018,Denholm_challenges_100pct_RE_system_USA_2021}, finally exceeding the cumulative \SI{1}{TW} mark in 2022 \cite{SolarPowerEurope_Global_Market_Outlook_2022}, the demand for more accurate power and energy forecasts by multiple stakeholders will increase at pace. In the design phase, PV system performance models in commercial software can (quite) accurately predict the power output of the system, provided that the user has selected the right components, data sources and made some important assumptions. Once the PV system has been financed and built, the physical components are expected to remain in place for the technical or financial lifetime of the plant. It is only then that the true interaction of the PV system components (modules, inverters, and mounting system) with the local weather and geographic conditions happens, including occurrences of rapidly changing irradiance and wind speed. Significant deviations between modelled and measured data have been observed among expert practitioners using the same software \cite{Stein_PV_performance_modelling_workshop_modelling_errors_2011,Moser_T13_Energy_Yield_benchmarking_PVSEC_2020}, highlighting the importance of calibrating or re-calculating system model coefficients from measured data.  

The contractual and practical consequences of this are that stakeholders are much better served by having the most accurate model of a system at their disposal at the fastest practical time resolution, rather than (only) a contractual model whose input parameters and time resolution are fixed (and perhaps too slow). One such component of the overall system is the thermal model, for which we show and investigate methods in this paper to achieve high accuracy at reasonable complexity, particularly at timescales ranging from seconds to minutes.

This paper presents a range of methods to improve the accuracy of equation-based thermal models of PV modules at second-to-minute timescales. We show that any explicit equation-based thermal model can be transformed from static to dynamic through the use of the exponential weighted mean (EWM) approach, while maintaining or improving that model's error metrics as time steps are shortened from hours to minutes or seconds. This gives a significant improvement versus known examples from the literature \cite{gofran_forced_convection_temperature_sensitivity_PVSC2019}. While the methods presented in this work can be used independently, their power is magnified when combined. The methods to improve existing and new thermal models and their understanding use: improved filtering of data, with the MBE as a proxy for steady-state radiation losses, and making these models dynamic through the use of the time constant $\tau$ for the EWM methodology, which are derived from the RC-equivalent thermal model of a PV module. The aim is to keep the resulting model complexity as low as possible, while obtaining high model accuracy. This is achieved by:
\begin{itemize}
	\item Using RC equivalent networks for conceptual understanding. This can serve to improve accuracy of backsheet-to-cell temperature corrections and understanding of wind direction effects;
	\item More robust filtering requirements for data, inspired by the RC conceptual model, resulting in more reliable coefficient determination;
	\item The determination of the equivalent thermal time constant $\tau$ of PV modules;  and,
	\item The use of $\tau$ for the exponential weighted mean (EWM) for irradiance and wind speed signals, making previously steady state models dynamic.

\end{itemize}

Ideal equation-based (thermal) models should:
\begin{itemize}
	\item Demonstrate high accuracy: low RMSE, MAE and MBE values;
	\item Have as few coefficients as possible for simplicity; and,
	\item Work with industry-standard signals (plane-of-array irradiance $G$, the ambient temperature $T_a$, module temperature $T_m$, and wind speed $WS$) for wide applicability.
\end{itemize}

This paper is organised as follows: we discuss a few noteworthy approaches for dynamic thermal models from the literature, highlighting strengths and weaknesses. We subsequently present the RC-equivalent thermal model of a PV module, which is used for conceptual understanding, to help determine the thermal time constant $\tau$, and to select the optimal filtering conditions to find model coefficients. We show that finding the optimal model coefficients are best achieved via multiple linear regression instead of simple linear regression. With $\tau$ known, it is then possible to calculate the Exponential Weighted Mean (EWM) irradiance and wind speed signals, and use these to make the (previously) steady-state thermal model dynamic. By correcting the Mean Bias Error (MBE) of the testing dataset (using it as a fixed radiation loss component), the thermal models are further improved. The filtering - EWM - MBE correction (FEM) methodology is applied on five thermal models and 24 datasets of varying time resolution: two thermal model variants we introduce (WM1 and WM2), which are compared against those of Ross \cite{Ross_thermal_model_1976}, King et al \cite{Sandia_King_PV_array_perf_model_2004} and Faiman \cite{Faiman_thermal_model_PiP_2008}. We then evaluate the thermal models along various dimensions and timescales, contextualising the model results against measured data.


\printglossary

\section{Literature review of dynamic thermal models} 

To model the temperature of a PV module from an energy balance perspective, radiation, convection (free and forced) and conduction as well as electrical power removal should be considered; see e.g. \cite{Mattei_PV_temperature_energy_balance_2006}. In practice, most explicit (empirical) thermal models disregard (variable) radiation thermal losses, as well as the electrical power removed from the module. Most empirical or data-driven thermal models therefore consider a \textit{residual} effect, i.e. the temperature of a module \textit{after} radiation and electrical power fluxes have been removed.

Physics-based approaches typically focus on coupled thermal-electrical models such as those by Tina \cite{Tina_Coupled_electrical_thermal_2010}, Goverde et al \cite{Goverde_coupled_electrical_thermal_energy_model_PVSEC_2014}, and Gu et al \cite{Gu_electrical_thermal_model_dynamic_2019}. These models arrive at a module temperature through a bottom-up physics-based approach, with the disadvantage of computational cost and complexity for wide implementation. These are better suited for attribution, i.e. to answer questions such as ``how large is the impact of radiation/convection/conduction/... on the module temperature?''. Importantly, the (relative) attribution differs significantly from bottom-up models to empirical equation-based models, in that coefficients describing the same effect (e.g. irradiance heating or convection cooling) are quite different. 

Lobera and Valkealahti \cite{Lobera_Valkealahti_Dynamic_thermal_model_2013} developed a dynamic thermal model from an energy balance approach, which they tested using \SI{1}{s} data over three months, achieving RMSE values of \SIrange{1.12}{1.61}{K}, depending on the time period considered in the dataset. In their work, the equation for the module temperature needs to be solved using Euler's method. While the authors claim it not to be computationally intensive, extensive work on finding the optimal coefficients from measured data was required.

A noteworthy equation-based dynamic model for PV module temperatures is by Veldhuis et al \cite{Veldhuis_Peters_EWM_model_2015}, which uses the exponential moving average of the module temperature through a recursive calculation on \SI{1}{min} data. It achieved average RMSE values of \SI{1.6}{K} for the modules tested in two locations, and gives a temperature lag (i.e. thermal time constant $\tau$) of \SI{17}{min}. The model consists of six parameters, which require irradiance (G), relative humidity (RH), wind speed (WS), module temperature ($T_m$) and ambient temperature ($T_a$) signals. The parameters are determined by minimising RMSE for a range of values, a process which hinders easy application for practical purposes. Peters and Nobre \cite{Peters_Thermal_Floating_PV_time_const_2022} apply this model for a floating PV system (RMSE \SI{2.3}{K}, $\tau =$ \SI{7}{min})  and compare this to rooftop PV (RMSE \SI{1.6}{K}, $\tau =$ \SI{35}{min}).
 
Prilliman et al \cite{Prilliman_Transient_thermal_model_2020} developed a transient weighted moving-average (or exponential weighted mean) thermal model, which takes the Sandia temperature model \cite{Sandia_King_PV_array_perf_model_2004}, and multiplies this with a combined exponential weighting coefficient \textit{P}, which itself depends on four coefficients $a_0$ to $a_3$. These coefficients can be determined from finite element analysis (FEA), which adds complexity for routine implementation. A fixed value for $\tau$ of \SI{20}{min} is employed for the calculation of the dynamic module temperature. For four sites at \SI{1}{min} time resolution, RMSE values \SIrange{2.0}{2.9}{K} are achieved, with mean bias error (MBE) results of \SI{\pm 0.8}{K} or smaller. 

Barry et al \cite{Barry_dynamic_PV_temperature_model_2020} proposed an extended form of Faiman's model \cite{Faiman_thermal_model_PiP_2008} as part of their dynamic modelling approach on \SI{1}{min} data, using a coefficient $u_3$ to multiply against the sky-ambient temperature difference. The determination of $u_3$ itself is not clear, yet it appears to depend on the sky temperature $T_{sky}$, which is obtained through measured data, using long-wave downward welling irradiance measured by a pyrgeometer. They show that the radiative cooling of PV modules can be linearised and used in the form of $u_3 \cdot \Delta T_{sky-ambient}$.

They determine thermal time constants for the three systems considered (System 1, 2A, and 2B) at \SIrange{500}{600}{s}, and achieve RMSE values for the three months of data as \SI{1.35}{K}, \SI{1.20}{K} and \SI{2.18}{K} respectively \cite{Barry_dynamic_PV_temperature_model_2020}.

While the aforementioned approaches permit the respective authors to obtain dynamic thermal models with significantly improved RMSE values at \SI{\sim 1}{min} time resolution, they often require non-standard measurement data, computationally intensive methods, or do not show a \textit{simple} yet robust methodology to obtain the required coefficients. It is not clear how to apply these methods at different time resolutions (e.g. \SI{1}{s} - \SI{10}{s} - \SI{1}{min}), and how the results vary accordingly.

\section{Methodology and data}

Fundamentally, we start by taking the (residual) RC-equivalent thermal network of a PV module shown in \Cref{Fig:RC equivalent model cross section}, similar to Armstrong et al \cite{Armstrong_Hurley_Thermal_Model_time_const_2010}, to re-evaluate and re-examine equation-based thermal model ``families''\footnote{The interested reader is also referred to \cite{Mattei_temperature_energy_balance_2006}, \cite{Skoplaki_simple_temp_correlation_2008}, \cite{Skoplaki_temperature_review_correlations_2009} for other explicit (empirical) models, similar in approach to the aforementioned three models, which are equally valid candidates for evaluation.} that have seen high uptake in the literature, namely those by Ross (\Cref{Eq:Ross_orig}) \cite{Ross_thermal_model_1976}, King et al \cite{Sandia_King_PV_array_perf_model_2004} (also known as the Sandia model: \Cref{Eq:King_orig}) and Faiman \cite{Faiman_thermal_model_PiP_2008} (\Cref{Eq:Faiman_orig}). In practice, Faiman's model is typically used in its simplified form (\Cref{Eq:Faiman_simplified}) \cite{Barykina_Sensitivity_Faiman_climates_2017}, as determining the optical efficiency $\eta_o$ requires additional effort, while the electrical efficiency $\eta_e$ varies as a function of the module's temperature. For notational and conceptual simplicity, and recognising that most models incorporate the ambient temperature $T_a$, the over-temperature $T_o$ defined by Kurnik et al \cite{kurnik_topic_outdoor_testing_PV_modules_overtemperature} and given in \Cref{Eq:Kurnik} will be used throughout this work.

\begin{align}
	T_o =&\: T_{m} - T_a \label{Eq:Kurnik} \\
	T_{o,Ross} =&\: k \cdot G \label{Eq:Ross_orig} \\
	T_{o,King} =&\: G \cdot e^{a+b\cdot WS} \label{Eq:King_orig} \\
	T_{o,Faiman,original} =&\: \frac{G}{ \frac{U_0}{\eta_o - \eta_e} + \frac{U_1}{\eta_o - \eta_e}\cdot WS}  \label{Eq:Faiman_orig}		\\
		T_{o,Faiman} =&\: \frac{G}{ U_0 + U_1\cdot WS}  \label{Eq:Faiman_simplified}
\end{align}

\begin{figure}[htb]
	\centering
	\includegraphics[width=\linewidth]{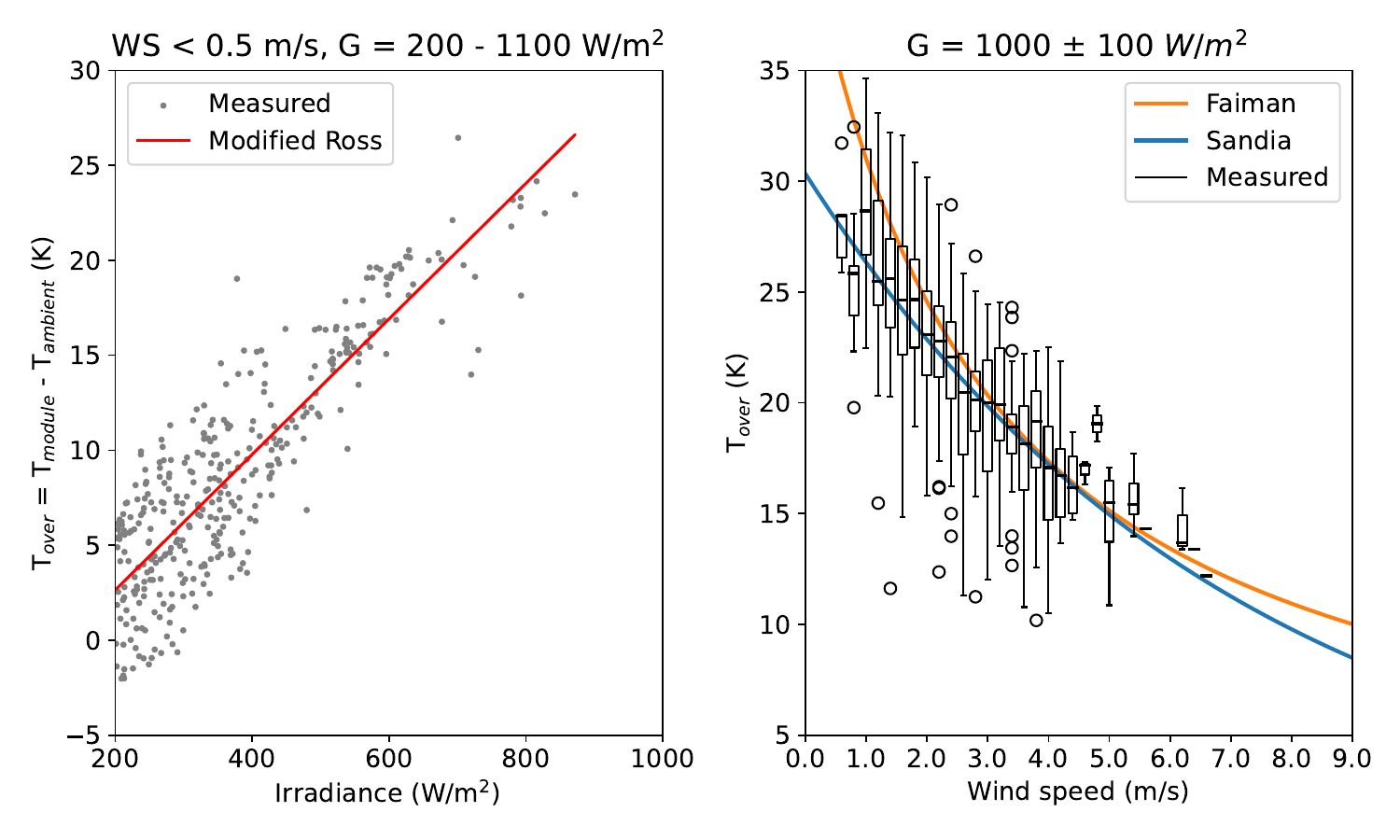}
	\caption{Motivation for examining the models by Ross, Sandia, and Faiman. The Ross (irradiance) coefficient $k$ reflects irradiance impacts (left subplot), whereas the Faiman and Sandia models approximate wind cooling (right subplot) quite well, although a divergence at low and high wind speeds is noticeable. Data filters are given for each subplot; data resampled to \SI{5}{min} averages.}
	\label{Fig:Model_motivation}
\end{figure} 

\subsection{Theoretical approach and conceptual RC model}

From electrical network theory, the dynamic behaviour of series-connected RC networks subject to step changes and the associated electric time constant $\tau_{el}$ is well understood. 
In outdoor conditions, PV modules and systems are subjected to a \textit{sequence} of irradiance and wind speed step changes of varying magnitude, which interact with the PV module and the mounting methods, thereby affecting the network equivalent thermal resistance and capacitance values. The module temperature evolves per time step, starting from the temperature attained in the previous time step. We will show that such a sequence of step changes of irradiance and wind speed is mathematically identical to using the exponential weighted mean (EWM) of those signals, with (near-) universal applicability to other explicit equation-based thermal models of PV modules and systems, making those models dynamic, rather than static\footnote{The approach by Prilliman et al \cite{Prilliman_Transient_thermal_model_2020} also allows static models to become dynamic.}. 

\begin{figure}[htb]
	\centering
	\includegraphics[width=\linewidth]{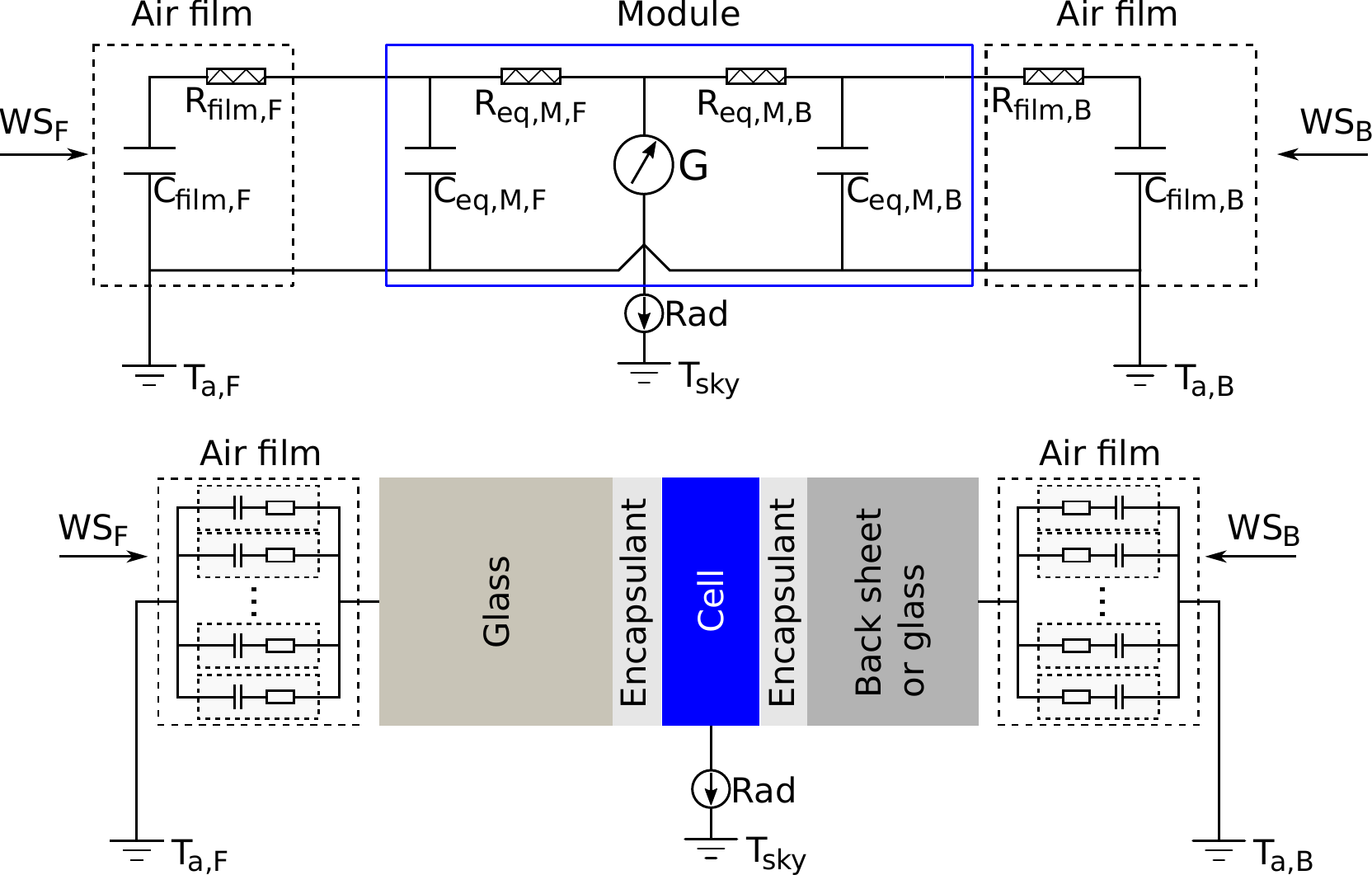}
	\caption{Conceptual RC model with simplified module cross section (not to scale). The ambient temperature and the wind speed on the front and back of the module may differ. The equivalent air films are in series to the glass or backsheet for convective heat transfer; the air film itself can be seen as composed of multiple air films in parallel, depending on wind speed. }
	\label{Fig:RC equivalent model cross section}
\end{figure}

The RC conceptual model in \Cref{Fig:RC equivalent model cross section} shows a ``standard'' PV module, with heat generation within the cell, which is dissipated towards the front and back surfaces through a series of RC networks; the front and back are thus in parallel to each other. Radiative heat losses are seen as constant. The air film on back and front of the module is an additional RC-equivalent network in series with each surface. Each (equivalent) air film can be seen as being composed of one or more air films in parallel to each other. The equivalent air film is impacted by the wind speed on that surface: 
\begin{itemize}
	\item At high wind speeds, the air film at the surface(s) of the module can be refreshed more often, reducing its equivalent thickness $L$. The equivalent air film resistance per unit area $r_{film,eq}$ is then minimal, while the air film capacitances per unit area add up, increasing $c_{film,eq}$. For $WS \approx \infty$, $r_{film} \approx 0$, $c_{film} \approx c_{film,max}$.
	\item For near-zero wind speeds, the equivalent values of $r_{film,eq}$ and $c_{film,eq}$ stem from the reduced number of air films in parallel: this increases $r_{film,eq}$, while it reduces $c_{film,eq}$. For $WS \approx 0$, $r_{film,eq} \approx r_{film,max}$, $c_{film,eq} \approx c_{film,min}$.

\end{itemize}
While the relative humidity affects the thermal capacity of air and therefore $c_{film,eq}$, it is not (yet) routinely measured for many PV systems. Given the lack of this data within the available datasets used here (see \Cref{Tab:data_sources}), the impact of the relative humidity is therefore disregarded in this work. 

The equivalent thermal resistance per unit area or R-value $r_{eq}$, equivalent thermal resistance $R_{eq}$, thermal capacitance per unit area or C-value $c_{eq}$,  thermal capacitance $C_{eq}$, and overall equivalent thermal time constant $\tau$ are defined as:
\begin{align}
	R_{eq} &= \frac{L}{\lambda \cdot A } &\bigg[\frac{K}{W}\bigg] \label{Eq:Req} \\
	r_{eq} &= R_{eq} \cdot A = \frac{L}{\lambda \cdot \cancel{A} } \cdot \cancel{A} = \frac{L}{\lambda} &\bigg[\frac{K}{W/m^2}\bigg] \label{Eq:req_spec}\\
	C_{eq} &= \rho \cdot A \cdot c_p \cdot L  &\bigg[\frac{J}{K}\bigg] \label{Eq:Ceq} \\
	c_{eq} &= \frac{C_{eq}}{A} = \frac{\rho \cdot \cancel{A} \cdot c_p \cdot L}{\cancel{A}} = \rho \cdot c_p \cdot L  &\bigg[\frac{J}{K \cdot m^2}\bigg]  \label{Eq:ceq_spec} \\
	\tau &= R_{eq} \cdot C_{eq} = r_{eq} \cdot c_{eq} = \frac{L^2 \cdot \rho \cdot c_p}{\lambda}   &[s] \label{Eq:tau_thermal_basic}
\end{align}
with $L$ the thickness of the material (\si{m}), $\lambda$ the thermal conductivity of the material (\si{W/(m\cdot K)}), $A$ the surface area (\si{m^2}), $\rho$ the material density (\si{kg/m^3}), and $c_p$ the specific heat capacity (\si{J/(kg\cdot K)}). The attentive reader will notice $\tau$ being expressed in seconds, in line with its electrical definition.

\Cref{Tab:R_C_tau_theoretical} gives theoretical values for a generic PV module, with an air film calculated to obtain $r_{eq} \approx$ $34.15\frac{K}{(1000 \cdot W/ m^2)}$, and an equivalent area $A_{eq}$ for the aluminium frame. The contribution to $r_{eq}$ by the aluminium frame is minimal, yet the C-value of the is not negligible, thereby affecting the module's thermal time constant: setting the thickness $L$ of the Al frame to zero, results in $\tau_0$~\SI{\approx 259}{s}. We denote $\tau_0$ as the value of $\tau$ at near-zero wind speeds. The values in \Cref{Tab:R_C_tau_theoretical} illustrate the large impact of the air films on the value of $r_{eq,total}$, whereas $r_{eq,module}$ is an order of magnitude smaller. This aligns with known properties of air, which is an excellent insulator (low thermal conductivity $\lambda$ or high thermal impedance ($1/\lambda$) \textit{and} relatively high thermal capacity $c_p$), \textit{provided the air remains in place}. Moreover, modifying the air film at the back of a PV module for on-roof systems to have an increased thickness $L =$ \SI{4.2}{mm}, gives $r_{eq} = 50\frac{mK}{(W\cdot m^2)}$, or expressed in terms of Faiman's simplified model, $U_0 = 20.0$, which hews closely to default values used in PVsyst ($U_c = \frac{15}{\eta_o - \eta_e} \approx \frac{15}{0.9-0.15} = 20$ for on-roof PV systems) \cite{PVsyst_thermal_manual_online}. 

From \Cref{Eq:tau_thermal_basic}, it is clear that $\tau \propto L^2$, so that the equivalent thickness of the module \textit{and} air film will affect $\tau$ strongly. For practical computational purposes using the pandas library in python \cite{pandas_package_mckinney2010data}, a constant value for $\tau$ must be used. 

\newcolumntype{C}{>{\centering\arraybackslash}p{0.12\linewidth}}
\newcolumntype{D}{>{\centering\arraybackslash}p{0.06\linewidth}}
\newcolumntype{E}{>{\centering\arraybackslash}p{0.075\linewidth}}

\begin{table*}[h!]
	\centering 
	\caption{Values for a representative glass-tedlar \textit{Smart PV} module installed at KU Leuven Technology Campus Ghent, rooftop array \cite{Herteleer_normalised_efficiency_2017}, with thermal data from \cite{Herteleer_PhD_thesis_2016} and the module manufacturer, Soltech.}
	\label{Tab:R_C_tau_theoretical}
	\begin {tabular}{CDDDDDDEED}%
	\toprule Layer&$L$ $[mm]$&$\lambda $ $\big [\frac {W}{m\cdot K}\big ]$&$\rho $ $\big [\frac {kg}{m^3}\big ]$&$c$ $\big [\frac {J}{kg \cdot K}\big ]$&$A_{eq}$ $[m^2]$&$m$ \quad $[kg]$&$r_{eq}$ $\big [\frac {mK}{W/m^2}\big ]$&$c_{eq}$ $\big [\frac {kJ}{K\cdot m^2}\big ]$&$\tau _0$ \quad $[s]$\\\hline %
	Air film$_{front}$&1.5&0.023&1.23&1000&1.6&0.00&65.22&0.0&0.1\\\hline %
	$\text {Al frame}_{front}$&2&237&2700&900&0.296&1.60&0.01&4.9&0.0\\%
	Glass&3.2&1.8&3000&500&1.6&15.36&1.78&4.8&8.5\\%
	EVA&0.5&0.35&960&2090&1.6&0.77&1.43&1.0&1.4\\%
	PV cells&0.1&148&2330&677&1.6&0.37&0.00&0.2&0.0\\%
	PV cells&0.1&148&2330&677&1.6&0.37&0.00&0.2&0.0\\%
	EVA&0.5&0.35&960&2090&1.6&0.77&1.43&1.0&1.4\\%
	Tedlar&0.3&0.2&1200&1250&1.6&0.58&1.50&0.5&0.7\\%
	$\text {Al frame}_{back}$&2&237&2700&900&0.296&1.60&0.01&4.9&0.0\\\hline %
	Air film$_{back}$&1.5&0.023&1.23&1000&1.6&0.00&65.22&0.0&0.1\\\hline %
	$\text {Total}_{front}$&&&&&&&3.22&10.8&34.8\\%
	$\text {Total}_{back}$&&&&&&&2.94&6.5&19.0\\\hline %
	$\text {Total}_{front+air}$&&&&&&&68.43&10.8&740.6\\%
	$\text {Total}_{back+air}$&&&&&&&68.16&6.5&441.2\\\hline %
	$\text {Total}$&&&&&&21.42&1.54&17.3&26.5\\%
	$\text {Total}_{air}$&&&&&&21.42&34.15&17.3&590.6\\\bottomrule %
	\end {tabular}%
\end{table*}

\subsection{Datasets used, filters and time resolutions}

\Cref{Tab:data_sources} gives an overview of the data used in this work. This data comes from four main sources: KUL own measured data \cite{Herteleer_normalised_efficiency_2017,KUL_Rooftop_array_1s_weather_dataset_2015_2016}, open data from NIST PV arrays \cite{NIST_PV_dataset}, US DOE Regional Test Center and NREL data shared via \cite{T13_data}, and data used by Barry et al for their dynamic thermal model \cite{Barry_dynamic_PV_temperature_model_2020,Barry_dyn_thermal_model_data_2020}. Of these, the KUL rooftop array and the NIST datasets stand out, as these have sub-minute time recording resolution data. These two sites (KUL rooftop and NIST Ground RTD 4) are then used to validate model quality at additional time resolutions (\SI{1}{min}, \SI{5}{min}, \SI{15}{min}, and \SI{1}{hour} averaged values).

The data treatment applied in this work to determine coefficients and testing the models is to split the dataset into a training dataset (weekdays) and testing dataset (weekend days). In this way, both the training and testing datasets see the same range of seasons. The resulting Key Performance Indicators (KPIs): RMSE, MAE, MBE are reported here for the testing dataset.

The following data selection filters are employed: 
\begin{itemize}
	\item All data:
	\subitem \SI{0}{m/s} $\leq WS \leq$ \SI{25}{m/s};
	\subitem \SI{-20}{\degreeCelsius} $\leq T_{a} \leq$ \SI{60}{\degreeCelsius};
	\item Night-time data (for module bias determination):  $T_o$ when $G<$ \SI{20}{W/m^2};
	\subitem \SI{-20}{\degreeCelsius} $\leq T_{m} \leq$ \SI{80}{\degreeCelsius} 
	\item Daytime values: $G>$ \SI{20}{W/m^2}; and,
	\subitem \SI{0}{\degreeCelsius} $\leq T_{m} \leq$ \SI{80}{\degreeCelsius} (avoid snow-covered modules).
\end{itemize}

\ctable[
caption = {Locations, data sources and key mounting characteristics of arrays used for thermal models. All sites used have plane-of-the-array irradiance data, $T_{a}$, $T_{m}$ (typically backsheet), and wind speed data, with very few having relative humidity and wind direction. More detailed system descriptions and metadata can be found in the cited references.},
label = {Tab:data_sources},
width = 0.99\textwidth,
pos = htb,
star,
]
{	p{0.155\textwidth} p{0.17\textwidth} p{0.157\textwidth} p{0.045\textwidth} p{0.13\textwidth} p{0.11\textwidth}  p{0.07\textwidth} }
{
	\tnote[a]{Angle not stated: Not available, not stated, or unknown. Raised = parking canopy}
	\tnote[b]{HAX: horizontal 1-axis tracker }
	\tnote[c]{H.U.: Heidelberg University}
	\tnote[d]{C = cell, BS = backsheet. RTD = Resistance temperature detector. Sensor name or number noted if multiple sensors available in dataset. 5x4 = Row 5, column 4 counting from top left, viewed from front (=centre cell of module).}
	\tnote[e]{Recording resolution as stored in the dataset. Some measured at \SI{1}{s} resolution, but stored data at \SI{10}{s} or \SI{1}{min} averages.}

}
{ \FL
	Organisation \& site name & Location & Data used & Time step\tmark[e] & Mounting, tilt \& azimuth\tmark[a] & Module sensor(s)\tmark[d] & Ref \ML
	KUL TC Ghent Rooftop  & Ghent, Belgium &  2015/05-2016/03  & \SI{1}{s} & Flat roof \SI{18}{\degree} S& Cell \& BS, PV052-5x4 &	\cite{Herteleer_normalised_efficiency_2017,KUL_Rooftop_array_1s_weather_dataset_2015_2016}	\NN
	KUL AgriPV & Dendermonde, Belgium &  2021/05-2022/03  & \SI{60}{s} & E-W \SI{\pm 50}{\degree} HAX & Front glass, RTD &	\cite{Willockx_Vertical_HAX_Bifi_PV_Grembergen_IEEE_PVSC_2022}	\NN
	NIST Ground  & Maryland, USA &   2016 Jan-Dec  & \SI{10}{s} & Ground \SI{20}{\degree} S & BS RTD4 &  \cite{NIST_PV_dataset,Boyd_NIST_PV_description}\NN
	NIST Ground  & Maryland, USA &   2016 Jan-Dec  & \SI{10}{s} & Ground \SI{20}{\degree} S & BS RTD8 & \cite{NIST_PV_dataset,Boyd_NIST_PV_description}\NN
	NIST Canopy W  & Maryland, USA &   2016 Jan-Dec  & \SI{10}{s} & Raised \SI{5}{\degree} W & BS RTD4 & \cite{NIST_PV_dataset,Boyd_NIST_PV_description}\NN 
	NIST Canopy E  & Maryland, USA &   2016 Jan-Dec  & \SI{10}{s} & Raised \SI{5}{\degree} E & BS RTD4 & \cite{NIST_PV_dataset,Boyd_NIST_PV_description}\NN 
	DOE c10hov6  & New Mexico, USA &   2016 Jan-Dec  & \SI{60}{s} & Ground \SI{35}{\degree} S & Backsheet & \cite{T13_data} \NN
	DOE t3pg1sv  & New Mexico, USA &  2016 Jan-Dec  & \SI{60}{s} & Ground \SI{35}{\degree} S& Backsheet & \cite{T13_data}\NN
	DOE luemkoy  & Vermont, USA &   2017/08-2018/05  & \SI{60}{s} & Ground \SI{35}{\degree} S& Backsheet & \cite{T13_data}\NN
	DOE lwcb907  & Vermont, USA & 2017/08-2018/05  & \SI{60}{s} & Ground \SI{35}{\degree} S& Backsheet & \cite{T13_data}\NN
	DOE wca0c5m  & Florida, USA &   2016 Jan-Dec  & \SI{60}{s} & Ground \SI{30}{\degree} S& Backsheet & \cite{T13_data}\NN
	DOE z0aygry  & Florida, USA &  2016 Jan-Dec  & \SI{60}{s} & Ground \SI{30}{\degree} S& Backsheet & \cite{T13_data}\NN 
	NREL Sanyo  & Colorado, USA &  2016 Jan-Dec  & \SI{60}{s} & Ground \SI{40}{\degree} S& BS \#2 & \cite{T13_data}\NN 
	
	H. U.\tmark[c] Syst 1  & Germany &  2018 Sep-Oct  & \SI{60}{s} & Ground & Backsheet & \cite{Barry_dyn_thermal_model_data_2020}\NN 
	H. U.\tmark[c] Syst 2A  & Germany &   2018 Sep-Oct  & \SI{60}{s} & Ground & Backsheet & \cite{Barry_dyn_thermal_model_data_2020} \NN 
	H. U.\tmark[c] Syst 2B  & Germany &   2019 Jul-Aug  & \SI{60}{s} & Sloped roof & Backsheet & \cite{Barry_dyn_thermal_model_data_2020} 
	\LL
	
}

\subsection{Improved filtering for robust and replicable coefficient determination}
In our proposed Filter - Exponential Weighted Mean - Mean Bias Correction (FEM) methodology, filtering is the first step.

Replicable determination of model coefficients is key for any model to be useful. While the explicit or implicit approaches as presented by the respective authors (Ross \cite{Ross_thermal_model_1976}, King et al \cite{Sandia_King_PV_array_perf_model_2004}, and Faiman \cite{Faiman_thermal_model_PiP_2008}) can be applied, their replicability when using (nearly) full-year datasets leave something to be desired\footnote{The authors were more concerned with rapid coefficient determination from e.g. one week's worth of data \cite{Faiman_thermal_model_PiP_2008}.}. 

If instead we link the coefficients to fundamentals as seen from the RC-equivalent model, improved filtering approaches can be identified.

\subsubsection{Determination of $r_{eq}$, $r_{M}$ and $r_{film}$}
The determination of the maximum equivalent R-value $r_{eq,max}$ is achieved by solving \Cref{Eq:Ross_orig}, \Cref{Eq:King_orig}, or \Cref{Eq:Faiman_simplified}\footnote{In the case of Faiman's model, occasional outliers for $U_0$ and $U_1$ were corrected respectively using the standard values for $U_0 = 25$ and $U_1 = 6.84$.}, by setting the wind speed equal to zero under steady-state conditions:
\begin{align}
	r_{eq,max} = \frac{T_o}{G}\bigg|_{WS = 0} \:&\bigg[\frac{K}{W/m^2}\bigg] \label{Eq:Rth_init}\\ 
	\Rightarrow r_{eq,max} = k = e^{a} = \frac{1}{U_{0}} \:&\bigg[\frac{K}{W/m^2}\bigg] \label{Eq:Rth_model_equivalence}
\end{align}
Steady-state conditions can be approximated by using averaged data, with intervals \SI{\geq 5}{min}. In practice, $r_{eq,max}$ is determined from a regression of $T_o$ versus $G$ at near-zero wind speeds (\SI{<0.5}{m/s}). By contrast, the minimum equivalent R-value $r_{eq,min}$ occurs at maximum wind speed ($WS \rightarrow \infty$):
\begin{align}
	r_{eq,min} &= \frac{T_o}{G}\bigg|_{WS \rightarrow \infty} \approx r_{M} \:\bigg[\frac{K}{W/m^2}\bigg] \label{Eq:Rth_init_min}\\ 
	\Rightarrow r_{eq,min} &\underbrace{=}_{WS \rightarrow \infty}  e^{a+b*WS} = \frac{1}{U_{0}+U_1\cdot WS}  \label{Eq:Rth_model_equivalence_min}
\end{align}

\begin{figure}[htb]
	\centering
	\includegraphics[width=0.9\linewidth]{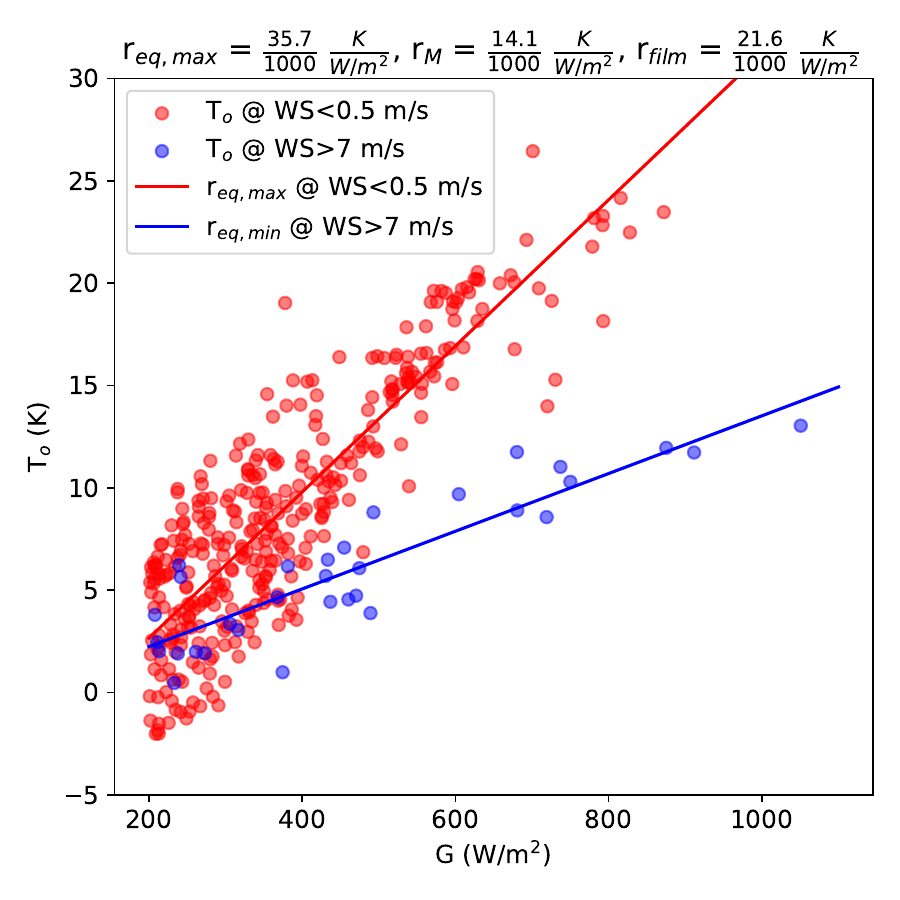}
	\caption{Example determination of $r_{eq}$, $r_{M}$ and $r_{film}$.}
	\label{Fig:R_th_eq Rth_M and R_film determination}
\end{figure}

To then determine $r_{film}$, it suffices to obtain the value of $r_{eq}$ at low and high wind speeds, using \Cref{Eq:Rth_eq_Rth_M_Rth_film}, as shown in \Cref{Fig:R_th_eq Rth_M and R_film determination}.
\begin{equation}
	r_{eq,tot} = \underbrace{r_{M}}_{r_{eq,min} @ WS \approx \infty} + 	\underbrace{r_{film}}_{\Delta r_{eq} = \underbrace{r_{eq,max}}_{@ WS \approx 0} - r_{eq,min} }  \label{Eq:Rth_eq_Rth_M_Rth_film}
\end{equation}

It must be noted however, that $r_M$ as calculated here, is the \textit{residual} and weather-impacted value of $r_M$, which will deviate from the physics-based calculations. 

Thus, \textit{when wind speed data is available}, filtering data at e.g. \SI{5}{min} time resolution for wind speeds as close as possible to zero, and maximum wind speeds, then gives the conditions through which to determine $r_{eq}$, by fitting a regression line to the respective filtered datasets. 

The temperature increase of PV modules due to irradiance, e.g. with $G = G_{STC}$, can then be calculated. A value\footnote{While numerically $r_{eq,max} = k_{Ross}$ has often been reported in the literature as e.g. $k = 0.035$, it helps to understand and communicate that such a module would be \SI{35}{K} hotter than ambient temperature with $G = G_{STC}$, which is achieved by the slightly more onerous notation we employ.} of $r_{eq,max} = \frac{35}{1000}\:\big[\frac{K}{W/m^2}\big] = 35\big[\frac{mK}{W/m^2}\big]$ thus results in a module's over-temperature $T_{o}$ versus ambient being equal to \SI{35}{K}, which can be determined both outdoors and indoors in a laboratory setting.

\begin{figure*}[h!]
	\centering
	\begin{subfigure}[b]{0.35\textwidth}
		\centering
		\includegraphics[width=\textwidth]{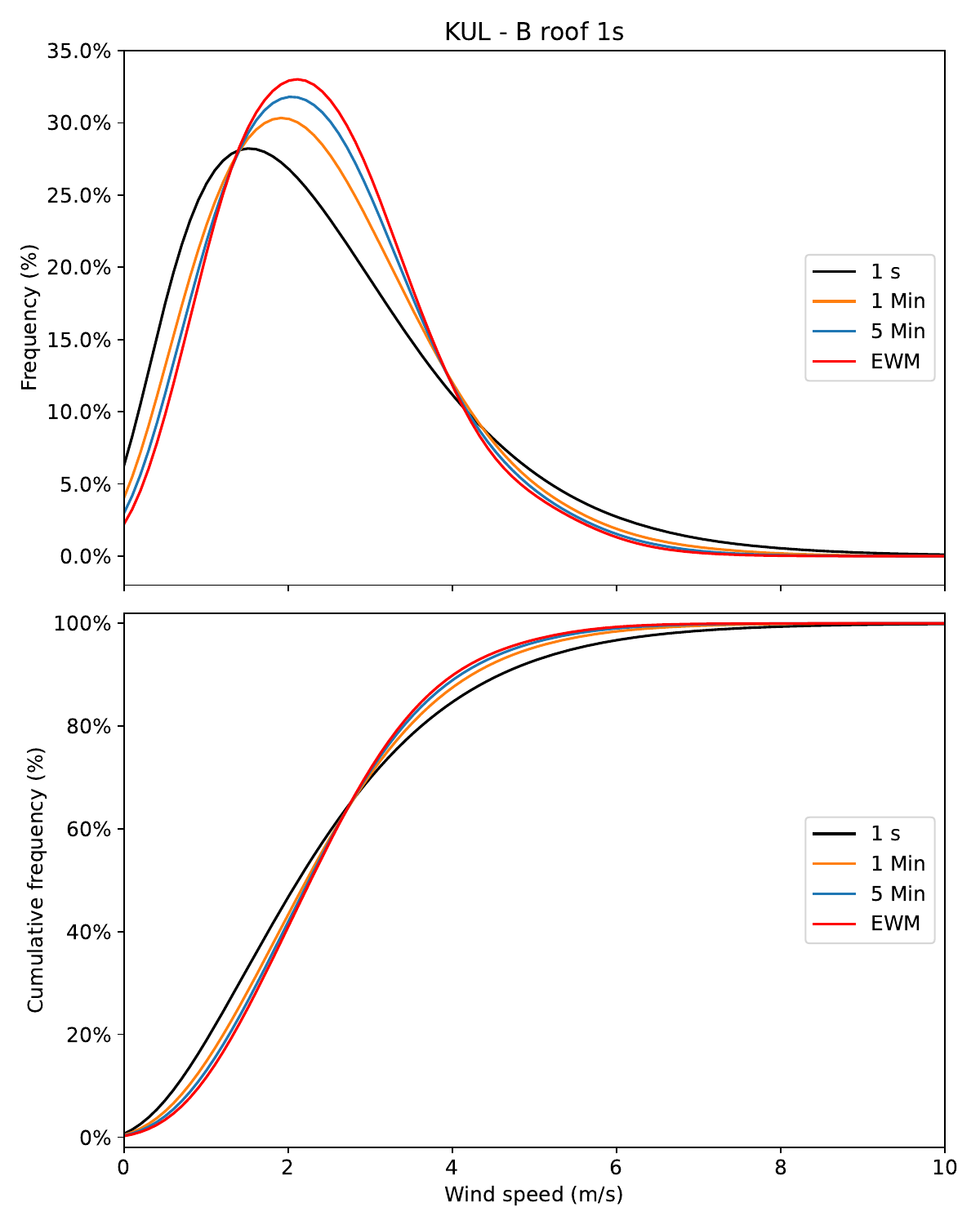} 
	\end{subfigure}
	\begin{subfigure}[b]{0.35\textwidth}
		\centering
		\includegraphics[width=\textwidth]{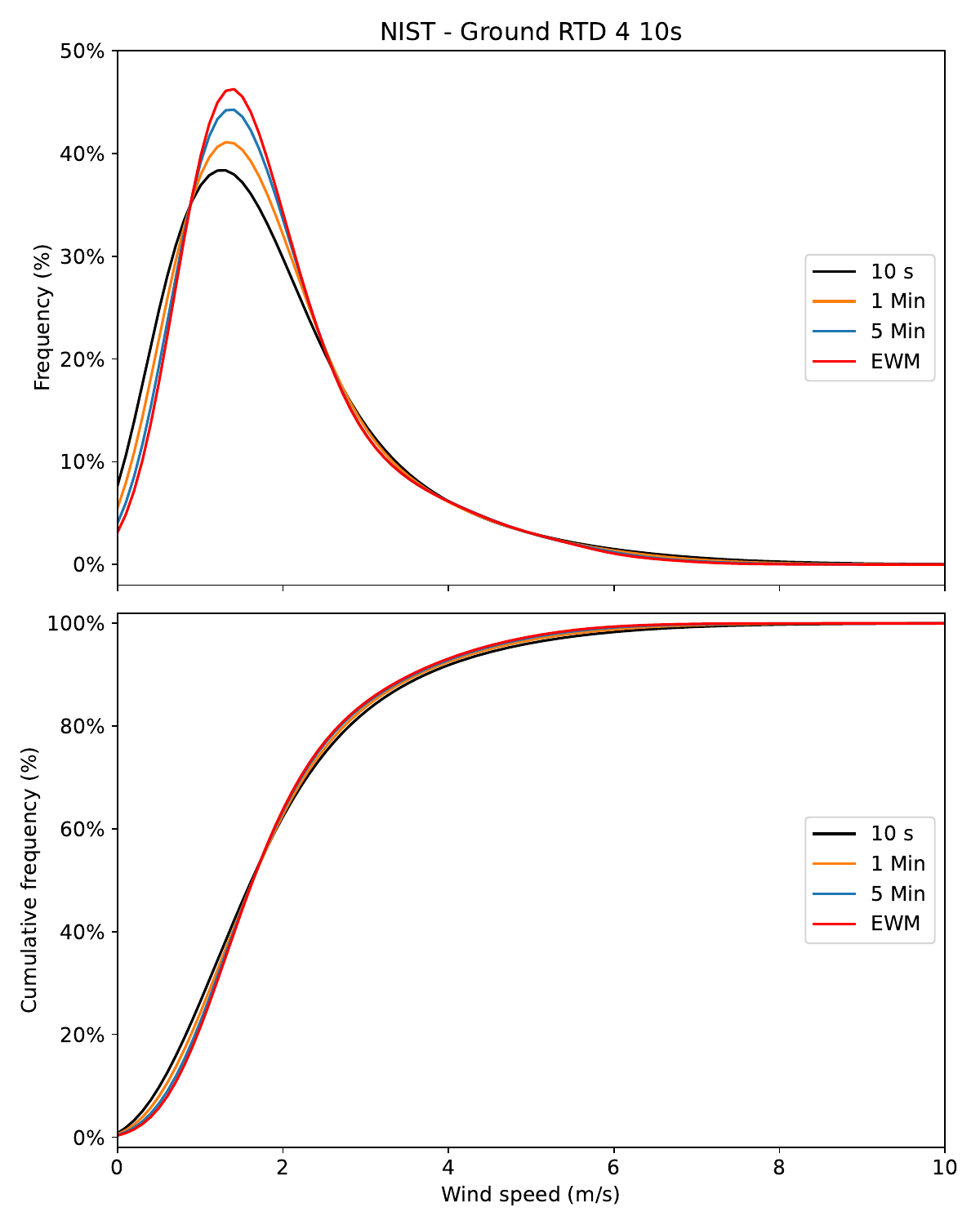} 
		
	\end{subfigure}
	\caption{Kernel density estimate distributions of wind speed values, depending on time resolution and data treatment, for KUL Ghent and NIST Ground arrays. These show the well-known Weibull distribution. For both sites, wind speeds above \SI{8}{m/s} are exceptionally rare, with \SI{90}{\%} of wind speeds below \SI{5}{m/s}. Note also the shift in the peak (most frequently occurring) wind speed for the different time resolution and treatment.}
	\label{Fig:WS_kdes_KUL_NIST}
\end{figure*}

The effective maximum wind speed to use in thermal models can be established from the frequency distribution of the wind speeds for a site. \Cref{Fig:WS_kdes_KUL_NIST} shows that a reasonable estimate for the maximum wind speed to use for thermal models is \SIrange{6}{8}{m/s} for \SI{5}{min} data. However, some sites may have very low wind speeds recorded, due to either the local conditions (e.g. nearby buildings or trees that block winds), or the placement of the wind sensor not being representative for the array.

\subsubsection{Determination of $c_{eq}$ from $\tau$}
The equivalent C-value $c_{eq}$ of a module cannot be determined directly from the measured data; instead, it is calculated from \Cref{Eq:tau_thermal_basic}. Similar to the determination of $r_{eq,max}$ and $r_{eq,min}$, the minimum and maximum equivalent C-value $c_{eq}$ require data from low and high wind speeds:

\begin{align}
	c_{eq,min} &= \frac{\tau_0}{r_{eq,max}}\bigg|_{WS = 0} \quad &\bigg[\frac{J}{K\cdot m^2}\bigg] \\
	c_{eq,max} &= \frac{\tau_0 \cdot e^{-WS/f}}{r_{eq,min}}\bigg|_{WS \rightarrow \infty} \quad &\bigg[\frac{J}{K\cdot m^2}\bigg]
\end{align}

\subsection{Proposed alternative thermal models}
\subsubsection{Wind Model 1}
We propose Wind Model 1 (WM1) as given in \Cref{Eq:Initial_heat_balance_eq}, which showcases the irradiance heating and convective cooling components: 

\begin{equation}
	T_o = \overbrace{k\cdot G \cdot \bigg(1}^{heating} \underbrace{- \big(1-e^\frac{-WS}{d }\big)}_{convective\: cooling} \bigg) = k\cdot G\cdot e^\frac{-WS}{d}
	\label{Eq:Initial_heat_balance_eq}
\end{equation}
A multivariate regression becomes possible for WM1, as $k = r_{eq,max}$ can be found when setting the wind speed equal to zero.

From \Cref{Eq:Rth_model_equivalence}, $k = e^a$ was identified. This allows the Sandia model to be rewritten, using $e^{a+b\cdot WS} = e^a\cdot e^{b\cdot WS}$ and $b = -\frac{1}{d}$, in the steady-state formulation:
\begin{equation}
	T_o =	G\cdot e^{a+b\cdot WS} = k \cdot G \cdot e^{-\frac{WS}{d}} \label{Eq:Sandia_WM1_equivalence_initial}
\end{equation}
As such, WM1 is functionally identical to the Sandia model when the same coefficients are used. In practice, WM1 will typically differ from the Sandia model, due to the different methods and philosophies used: multivariate or two-step linear regression for WM1 for low wind ($k = r_{eq,max}$) and $d$ for high irradiance and varying wind speeds, versus a single linear regression for the Sandia Model for all irradiance and wind conditions. 

Note also that \Cref{Eq:Initial_heat_balance_eq} with $WS = 0$ reduces to Ross's model formulation, thus giving a mathematical bridge between the different thermal model forms of Ross, Sandia, and WM1. By extension, this applies also to WM2 (see below) and Faiman's model. Driesse et al show that model parameters can be translated for the models of Faiman, King et al (Sandia), PVsyst, and the System Advisory Model NOCT approach, albeit using a different calculation route \cite{Driesse_PV_Module_Model_Equivalence_Translation}.

\subsubsection{Wind Model 2}

The second thermal wind model (WM2) that we propose also employs $k$. WM2 is given in its simplest form as

\begin{equation}
	T_o =  G \cdot \big(k - k_W\cdot WS\big|_{WS\leq 8} \big) = G \cdot k - G \cdot \underbrace{k_W\cdot WS\big|_{WS\leq 8}}_{ \frac{1}{h} } \label{Eq:WM2_wo_EWM}
\end{equation}

which has the wind speed clipped at \SI{8}{m/s}, informed by the knowledge from \Cref{Fig:Model_motivation,Fig:WS_kdes_KUL_NIST}. (This avoids $T_o$ becoming negative at very high wind speeds.) The convective heat transfer coefficient $h = \frac{1}{k_W\cdot WS} \quad \big[\frac{W}{m^2\cdot K}\big]$ is seen here as being composed of the wind speed and the coefficient $k_W$.  WM2 is essentially a more general form of thermal models in the literature \cite{Skoplaki_temperature_review_correlations_2009} of the form $T_M = T_a + k\cdot G - k_W\cdot WS + c$, albeit with a dynamically\footnote{It is also possible to calculate and use $1/h_{avg}$ as the annual mean of $1/h = G\cdot k_W$. $h_{fixed}$ calculated thus is similar to values reported in ref [49] from \cite{Skoplaki_temperature_review_correlations_2009}.} varying convection coefficient. \Cref{Eq:WM2_wo_EWM} reflects that higher convection losses are driven by higher irradiance, i.e. that the module must first gain heat via irradiance, before the wind can remove a larger amount of heat under high irradiance conditions.

\begin{figure*}[h!]
\centering
\includegraphics[width=0.8\textwidth]{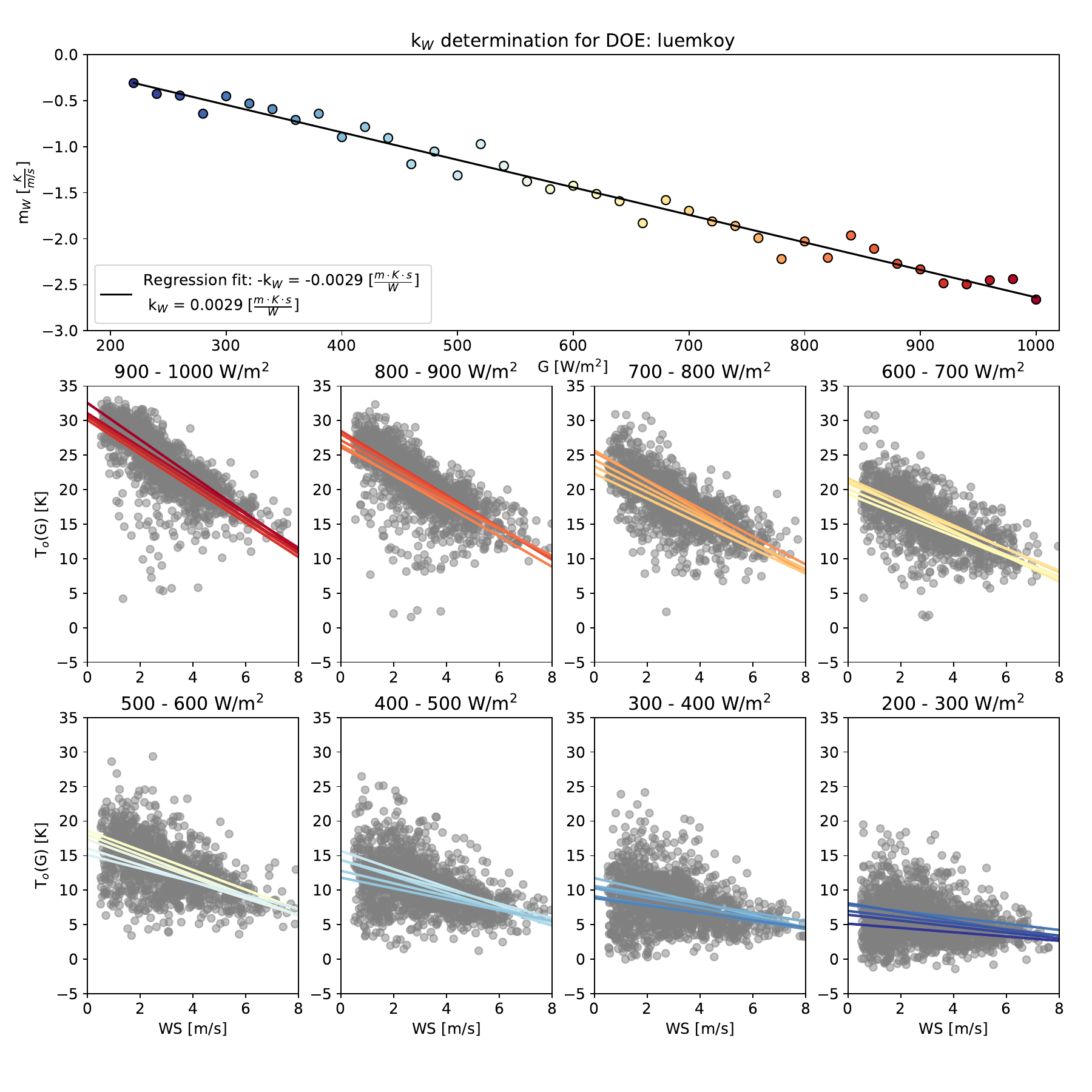}
\caption{Determination of the convection coefficient $k_W$ for WM2, for the DOE system luemkoy. Each coloured dot in the top pane is the respective regression line coefficient found at that \SI{20}{W\per m^2} irradiance bin. Due to the definition in \Cref{Eq:WM2_wo_EWM}, $k_W$ is typically a positive coefficient.}
\label{Fig:WM2_k_W_coeff_determination} 
\end{figure*}

The irradiance convection coefficient $k_W$ is calculated from a linear regression of linear regression coefficients, shown in \Cref{Fig:WM2_k_W_coeff_determination}:
\begin{itemize}
	\item For each irradiance bin of \SI{20}{W/m^2} width, for irradiances between \SIrange{200}{1000}{W/m^2}, find the regression coefficient $m_{W}(G)$ of $T_o$ versus wind speeds (\SIrange{0.5}{8}{m/s}).
	\item Fit a regression line of the $m_{W}(G)$ points versus irradiance $G$ (\SIrange{1000}{200}{W/m^2}). The resulting linear regression coefficient is $k_W$.
\end{itemize}

\subsection{Determination of the thermal time constant $\tau$}\label{Sec:determination_tau_thermal}
\begin{figure*}[h!]
	\centering
	\includegraphics[width=0.8\textwidth]{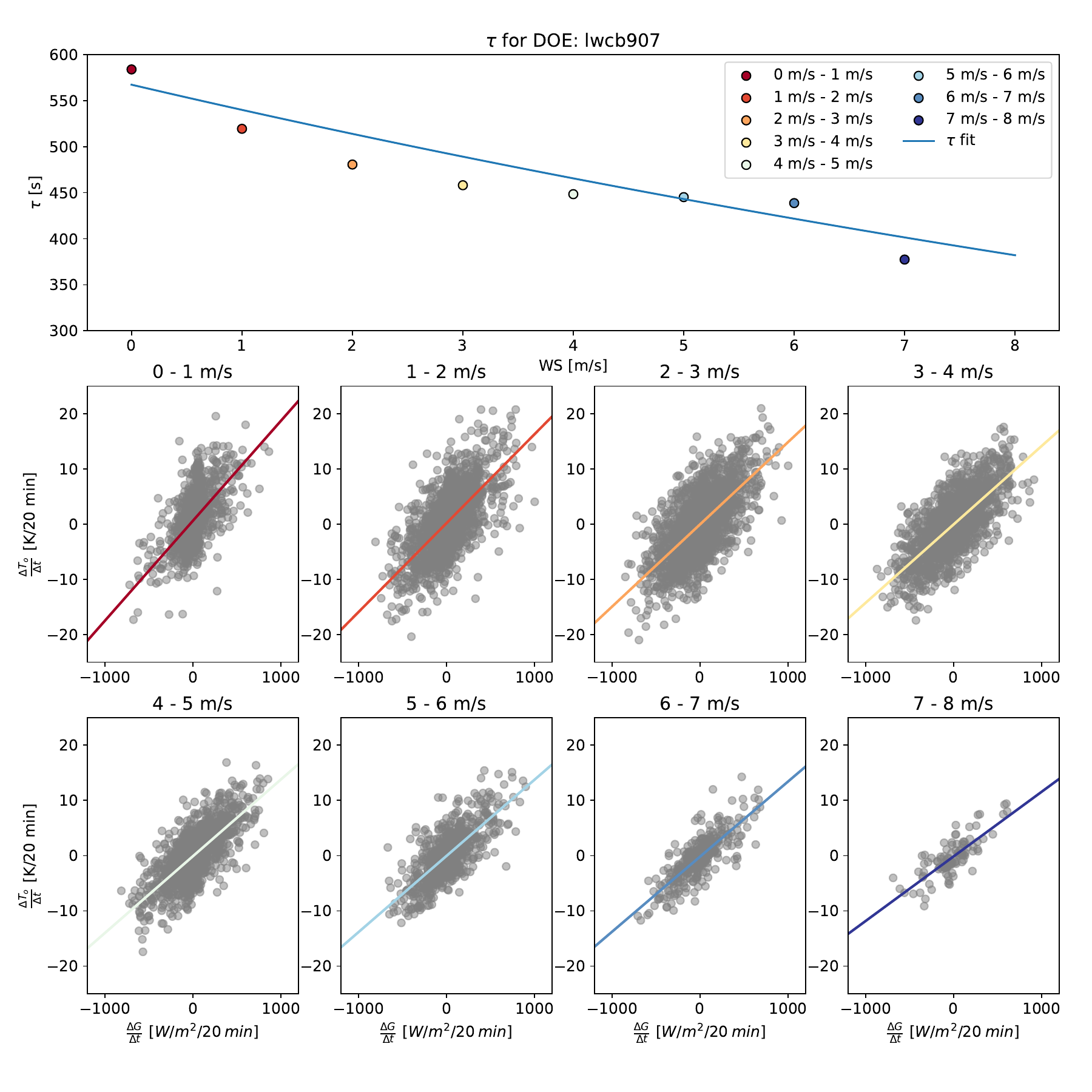}
	\caption{Determination of $\tau$ for the DOE system lwcb907. Each coloured dot in the top pane is the respective regression line coefficient found at that wind speed bin.}
	\label{Fig:tau_determination_regressions} 
\end{figure*}

The thermal time constant $\tau$ can be identified through \textit{sustained} step changes of both irradiance and the over-temperature. As the wind speed also affects $\tau$, multiple linear regressions for wind speed bins (\SIrange{0}{1}{m/s}, \SIrange{1}{2}{m/s}, ...) with minimal variation (i.e. avoiding wind gusts and lulls) need to be done, similar to how the convection coefficient $k_W$ for WM2 is calculated. An illustrative image is given in \Cref{Fig:tau_determination_regressions}.

For each wind speed bin:
\begin{itemize}
	\item \textbf{Filter}: Resample $T_o$ and $G$ to \SI{5}{min} averages:
	\subitem Daylight and sufficient module heating: $G>200\:W/m^2$, 
	\subitem Wind speed bin: $WS_{min,bin}<WS<WS_{max,bin}$, 
	\subitem Ensure limited wind speed variation: $|\Delta WS_{bin}| < \sigma_{WS,\text{full dataset}}$.
	\item \textbf{Find step changes}: Determine the difference for $T_{o,5min}$ and $G_{5min}$ between four \SI{5}{min} intervals, i.e. \SI{20}{min} apart\footnote{From the data, this appears to be the longest, most frequent, duration of low-high-low or high-low-high irradiance and $T_o$ step change sequences.} \textit{for each wind speed bin of} \SI{1}{m/s} width, from \SIrange{0}{8}{m/s}.
	\item \textbf{$\tau$ per wind speed bin}: determine the slope of the regression line of $\frac{\Delta T_{o}}{\Delta t}$ versus $\frac{\Delta G}{\Delta t}$, for a total time interval $\Delta t = $\SI{20}{min}. The aim here is to isolate the value of $\tau$ for each wind speed, which rests on using the formulation of WM1.
	$\tau (WS)$ is then:
	\begin{align}
		\tau(WS) &= \frac{\Delta T_{o} / \Delta t}{\Delta G / \Delta t} / r_{eq,max}  \cdot \Delta t \nonumber \\
		&= \frac{r_{eq}\cdot (\Delta G\cdot e^{-WS/d})/ \Delta t}{ \Delta G / \Delta t } / r_{eq,max}  \cdot \Delta t \nonumber \\
		&= \frac{( \cancel{\Delta G} \cdot e^{-WS/d}) / \cancel{\Delta t} }{ \cancel{\Delta G} / \cancel{\Delta t} } \cdot \underbrace{\frac{r_{eq} \cdot \Delta t}{r_{eq,max}}}_{\tau_0}  		\nonumber \\
		&= \tau_0 \cdot e^{-WS/d}   \equiv \tau_0 \cdot e^{-WS/f} \: [s].
	\end{align}
	Note that the wind speed coefficient $d$ from WM1 is different from the wind speed coefficient $f$, as the latter is determined for step changes, whereas $d$ is found in steady-state (-like) conditions. Generally, $f$ has a larger magnitude than $d$ (i.e. smaller effect), which corresponds to step changes in wind speed having a smaller effect than a cumulative consistent wind speed. 
\end{itemize}

\begin{figure*}[htb]
	\centering
	\includegraphics[width=1\linewidth]{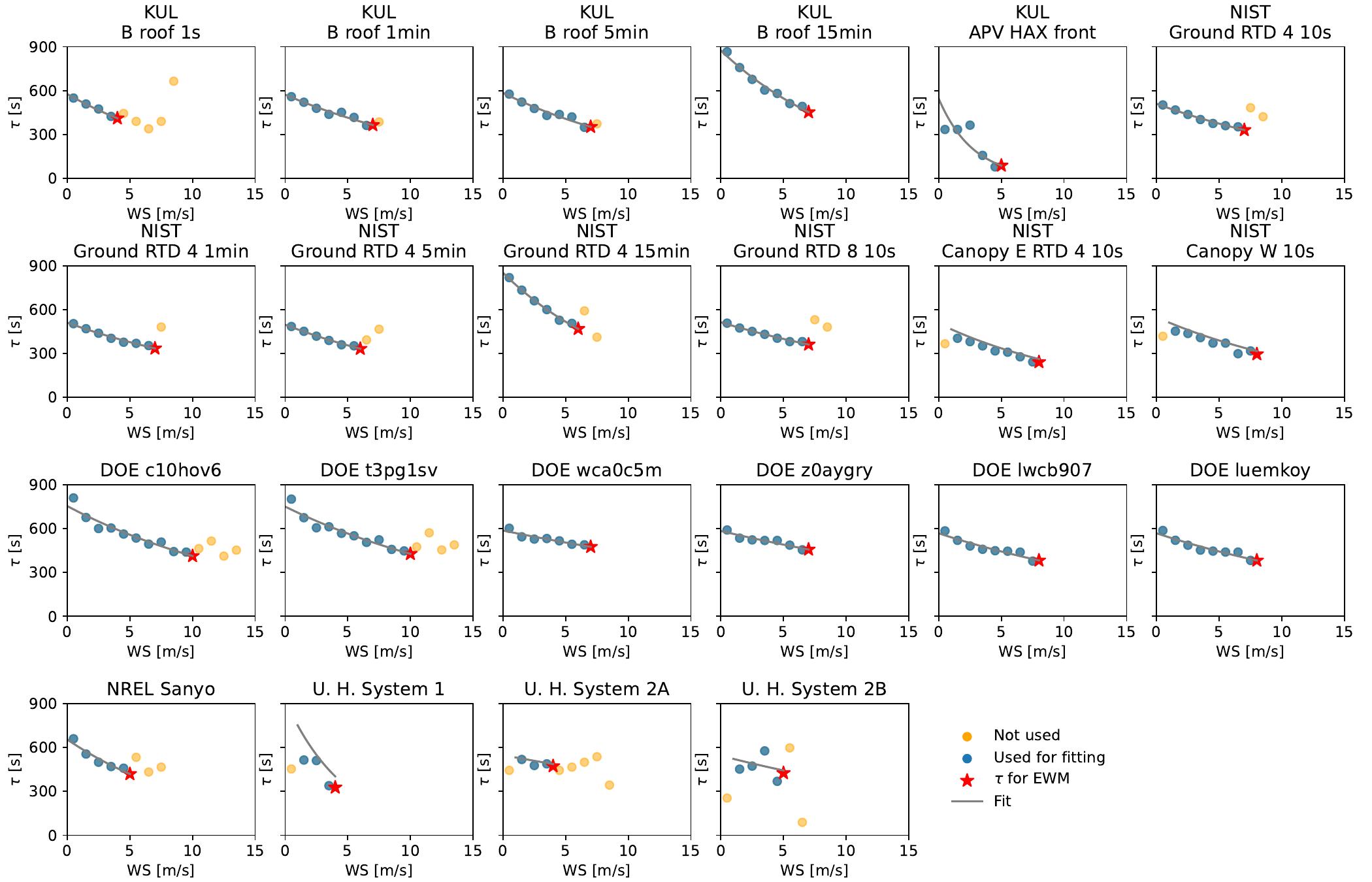}
	\caption{Values found for $\tau$ for wind speed bins for all datasets \SI{<1}{h}, with the excluded data points indicated, as well as the value of $\tau$ used for the EWM.}
	\label{Fig:tau_fitting_all_datasets}
\end{figure*}

The value of $\tau$ \textit{typically} decreases for increasing wind speed, as the module reaches its ``steady state'' situation more rapidly \cite{Armstrong_Hurley_Thermal_Model_time_const_2010}. This is also the basis for the equivalent wind film resistance $r_{film}$ to be variable, as it encompasses the interaction of module mounting and wind access to the module surface(s). For most sites, an exponential decrease of $\tau$ can be observed in \Cref{Fig:tau_fitting_all_datasets}. This relationship may break down for high wind speeds when it becomes highly turbulent (air films poorly or only partially refreshed over the module surface) or at very low wind speeds (insufficient or poor forced convection) at the module surface. Wind direction will thus affect $\tau$ and its determination, yet even in the absence of wind direction data, a serviceable value of $\tau$ can be found. 

With the knowledge that $\tau$ is not constant, the question arises as to which value results in lowest model errors when employed for the EWM calculation. Our experience indicates that $\tau$ needs to be sufficiently low to accurately capture faster irradiance and wind speed changes. This informs then that the value of $\tau$ used for the EWM approach is found as the last wind speed bin where $\tau$ is still decreasing monotonically, from a maximum value of $\tau_0$ found at \SIrange{0}{2}{m/s}, with an error margin of \SI{20}{s}, as shown in \Cref{Fig:tau_fitting_all_datasets}. The assumption of a purely monotonic decrease of $\tau$ versus the wind speed does not hold true for all sites. In particular for the systems (1, 2A and 2B) from \cite{Barry_dynamic_PV_temperature_model_2020}, it appears from the photographs that the wind direction and local wind barriers likely play a large role in module cooling, and therefore the determination of $\tau$. Similarly, the KUL single-axis tracker module sees a very aggressive decrease in $\tau$, which may be due to the tracker adjusting its position throughout the day, and thus seeing higher effective wind flow over the module, compared to a fixed orientation. 

\newcolumntype{F}{>{\centering\arraybackslash}p{0.21\linewidth}}
\newcolumntype{G}{>{\centering\arraybackslash}p{0.0425\linewidth}}
\newcolumntype{H}{>{\centering\arraybackslash}p{0.0525\linewidth}}
\newcolumntype{I}{>{\centering\arraybackslash}p{0.0325\linewidth}}

\Cref{Tab:r_c_tau_results} gives an average value and standard deviation $\sigma$ for $\tau$ for \SI{1}{s} to \SI{5}{min} data as \SI{364 \pm 90}{s} or \SI{6 \pm 1.5}{min} when the single-axis tracker data is included, and \SI{379 \pm 60}{s} or \SI{6.3 \pm 1}{min} when the single-axis tracker is excluded (i.e. for fixed mounting only). These results are broadly in line with the literature, with \cite{Armstrong_Hurley_Thermal_Model_time_const_2010} giving $\tau = $~\SI{6}{min} at $WS =$~\SI{2.14}{m/s}, $\tau = $~\SI{7}{min} for floating PV \cite{Peters_Thermal_Floating_PV_time_const_2022}, and for System 1, 2A, and 2B at \SI{10}{min}, \SI{8.5}{min}, and \SI{9}{min} respectively \cite{Barry_dynamic_PV_temperature_model_2020}.

\begin{table*}[h!]
	\centering 
	\caption{Measured-calculated data for the different datasets. $f$ is to calculate $\tau(WS)$, using $\tau_0$. The value of $\tau$ is used for the EWM approach within the FEM methodology in this work. The DOE data is separated by geographical sites, with two arrays per site. Contrast $\tau$ found here with values calculated by Barry et al for U. H. systems 1, 2A, and 2B at \SI{589}{s}, \SI{509}{s}, and \SI{547}{s} respectively \cite{Barry_dynamic_PV_temperature_model_2020}.}
	\label{Tab:r_c_tau_results}
	\begin {tabular}{HFGGGGGGIIII}%
	\toprule Org.&Site&$r_{eq,max}$ $\big [\frac {mK}{W/m^2}\big ]$&$r_M$ $\big [\frac {mK}{W/m^2}\big ]$&$r_{film}$ $\big [\frac {mK}{W/m^2}\big ]$&$c_{eq,max}$ $\big [\frac {kJ}{K\cdot m^2}\big ]$&$c_M$ $\big [\frac {kJ}{K\cdot m^2}\big ]$&$c_{film}$ $\big [\frac {kJ}{K\cdot m^2}\big ]$&$\tau _0$ \quad $[s]$&$\tau $ \quad $[s]$&$\Delta t$ \quad $[s]$&$f$ \quad $\big [\frac {s}{m}\big ]$\\\hline %
	KUL&B roof 1s&35.4&15.3&20.1&37.7&26.9&10.8&576&410&1&11.8\\%
	KUL&B roof 1min&34.8&15.3&19.5&37.4&23.8&13.6&572&364&60&15.53\\%
	KUL&B roof 5min&35.4&15.6&19.8&37.5&22.6&14.9&584&352&300&13.82\\%
	KUL&B roof 15min&31.2&16.3&14.9&53.7&27.8&25.9&874&453&900&10.64\\%
	KUL&B roof 1h&31.4&17.0&14.4&211.8&-&-&-&-&3600&-\\\hline %
	KUL&APV HAX front&12.2&5.3&6.9&102.8&16.6&86.2&545&87&60&2.74\\\hline %
	NIST&Ground RTD 4 10s&31.3&14.0&17.3&36.4&23.6&12.8&510&330&10&16.16\\%
	NIST&Ground RTD 4 1min&31.2&14.0&17.2&36.5&23.9&12.6&510&334&60&16.56\\%
	NIST&Ground RTD 4 5min&32.7&14.0&18.7&35.5&23.7&11.8&497&332&300&14.91\\%
	NIST&Ground RTD 4 15min&28.0&15.1&12.9&56.4&31.0&25.4&851&468&900&10.02\\%
	NIST&Ground RTD 4 1h&28.2&15.3&12.9&235.3&-&-&-&-&3600&-\\\hline %
	NIST&Ground RTD 8 10s&36.2&16.8&19.4&30.5&21.5&9.0&513&361&10&19.98\\%
	NIST&Canopy E RTD 4 10s&41.6&17.2&24.4&27.1&14.0&13.1&465&241&10&12.17\\%
	NIST&Canopy W 10s&38.2&18.5&19.7&27.6&16.0&11.6&511&295&10&14.57\\\hline %
	DOE&c10hov6&35.0&18.5&16.5&40.7&22.3&18.4&753&412&60&16.57\\%
	DOE&t3pg1sv&34.9&18.8&16.1&39.9&22.7&17.2&750&426&60&17.73\\\hline %
	DOE&wca0c5m&30.7&18.6&12.1&31.5&25.5&6.0&586&474&60&33.14\\%
	DOE&z0aygry&30.1&18.0&12.1&32.4&25.4&7.0&583&456&60&28.63\\\hline %
	DOE&lwcb907&37.2&19.1&18.1&29.7&20.0&9.7&567&382&60&20.24\\%
	DOE&luemkoy&35.8&17.7&18.1&32.1&21.6&10.5&567&381&60&20.18\\\hline %
	NREL&Sanyo&32.2&16.3&15.9&40.2&25.7&14.5&655&418&60&11.17\\\hline %
	U. H.&System 1&31.5&19.9&11.6&37.8&16.4&21.4&751&326&60&4.8\\%
	U. H.&System 2A&32.2&16.4&15.8&32.4&28.8&3.6&532&472&60&33.76\\%
	U. H.&System 2B&23.1&16.4&6.7&31.8&25.9&5.9&522&425&60&24.33\\\hline %
	Average&$\Delta t<900 \: s$&32.6&16.3&16.3&37.9&22.3&15.5&577&364&-&17.44\\\bottomrule %
	\end {tabular}%
	
\end{table*}

While not shown, most sites will see the KPIs (RMSE, MAE and MBE) vary moderately when $\tau$ differs from the value calculated using the approach shown in this work by \SIrange{0.5}{1}{min}, as long as $\tau$ is broadly in line with the expected value for that site and mounting conditions. As such, a value for $\tau$ in the range of \SIrange{5}{7}{min} can be safely used for most fixed-mount PV systems with open back. Further research on other mounting conditions, including tracking systems and on-roof arrays with reduced air gaps, is warranted.

\subsection{The exponential weighted mean}
The exponential weighted mean (EWM) is the second step in the FEM methodology.

For simplicity, assuming a situation with zero wind at steady state with high irradiance, sees the over-temperature (=voltage) at its maximum value. If the irradiance (=current) drops precipitously (e.g. cloud moving in front of the sun), this is the equivalent of opening a switch, with the thermal capacitor discharging its stored energy through the resistor $r_{eq}$. Over a time step $\Delta t$, the temperature drop of the module is mediated by:
\begin{align}
	\Delta T_o (\Delta t) &= \underbrace{\Delta G(\Delta t) \cdot r_{eq}}_{\Delta V} \cdot e^{-\Delta t/\tau} \\
	T_o(t + \Delta t) &= T_o(t) + \Delta T_o (\Delta t)
\end{align}
 
As such, it is evident that the temperature of a PV module depends on what happened in previous time steps. However, the impact of the past on the present decreases exponentially, as each time step is then subject to a change mediated by $1-e^{-\Delta t/\tau}$, i.e. a sequence of step changes of duration $\Delta t$ subject to an irradiance step change $\Delta G$ (positive, zero, or negative). This is the definition of an exponential weighted mean (EWM), which uses the smoothing parameter $\alpha$:
\begin{equation}
	\alpha = 1-e^{-\Delta t/\tau} \text{ with } 0\leq \alpha \leq 1
\end{equation}
The advantage of using $\alpha$ is that it adapts to a changing time resolution, so that the appropriately scaled effect is obtained for both \SI{1}{s} data and \SI{5}{min} data. The span is the number of time steps $\Delta t$ used for the calculation of the EWM, and relates to the smoothing parameter $\alpha$ (and the time constant $\tau$):
\begin{equation}
	span = \frac{2}{\alpha}-1 = \frac{2}{1-e^{-\Delta t/\tau}}-1 \underbrace{\simeq}_{\Delta t \leq \tau} 2\cdot \tau
\end{equation}

For a value of $\tau =$ \SI{300}{s} (\SI{5}{min}), the EWM will thus incorporate the impact of irradiance up to \SI{10}{min} in the past. This also suggests that this would be the time taken for the model to ``catch up'' to measured data in special conditions (early mornings, after heavy rain events), where the model error inevitably would be larger than during the rest of the day, as the EWM calculation has insufficient data, possibly combined with effects such as evaporative cooling after a rain event. 

The exponentially weighted mean of the irradiance is then:
\begin{equation}
		G_{EWM} =\frac{ \sum_{i=0}^{t=span}G_{i}\cdot w_i}{ \sum_{i=0}^{t=span}w_i} =\frac{ \sum_{i=0}^{t=span}G_{i}\cdot (1-\alpha)^i}{ \sum_{i=0}^{t=span}(1-\alpha)^i}
\end{equation}

with weights $w_i$ at time step $i$ given by:

\begin{equation}
	w_i = (1-\alpha)^i.
\end{equation}

Similarly for a (sustained) step change in wind speed, an equivalent effect will be seen on PV modules. Compared to irradiance, the wind speed is significantly more variable (less persistent), and thus makes identifying wind speed step change effects more challenging.  

With the above knowledge, the five models discussed in this work can be made dynamic through the use of EWM, achieving an improved RMSE and MAE at minute-to-second timescales. The EWM is applied to both the irradiance and wind speed signals. In python and pandas, making WM1 dynamic from its static form is as follows:
\begin{lstlisting}
To_WM1 = k*df[`G']*np.exp(-WS/d)
alpha_EWM = 1-np.exp(-delta_t/tau)
To_WM1_EWM = k*df[`G'].ewm(alpha = alpha_EWM).mean()*np.exp(-WS.ewm(alpha = alpha_EWM).mean()/d)
\end{lstlisting}
With knowledge of the time resolution $\Delta t$ and $\tau$, it is thus a simple change for a model to become dynamic, which can be applied even in the absence of measured data to calibrate coefficients as part of the FEM approach.

\subsection{Mean bias error correction}

The last step in the Filter-EWM-Mean bias error correction (FEM) methodology is using the mean bias error (MBE) from the training dataset, as given in \Cref{Eq:To_FEM}. A bias is often observed in thermal model data. Whereas Veldhuis et al \cite{Veldhuis_Peters_EWM_model_2015} calculate this as the mean night-time radiation bias $ = T_{o,night,avg}$, this appears to give (slightly) larger errors (RMSE, MAE, and MBE) than when using the MBE correction. Another noteworthy approach by Driesse et al \cite{Driesse_Incorporating_Radiative_Losses_Improving_Thermal_Models_2022} could also be explored. Conceptually, the MBE correction can be seen as the fixed radiative heat loss from the module, as shown in \Cref{Fig:RC equivalent model cross section}. To illustrate the full FEM approach, WM1$_{FEM}$ is written out in full in \Cref{Eq:To_FEM_WM1}, with the same method applied to all models in this work.

\begin{align}
	T_{o,FEM} = &T_{o,EWM,test} \nonumber \\  &+ MBE_{train} \label{Eq:To_FEM} \\
	T_{o,FEM,WM1} = &k_{test}\cdot G_{EWM}\cdot e^{-WS_{EWM}/d_{test}} \nonumber \\
	&+ MBE_{train} \label{Eq:To_FEM_WM1}
\end{align}

\section{Results \& Discussion}

\subsection{FEM versus standard approach on different timescales}\label{Subsec:FEM_vs_std_timescales}
\begin{figure*}[htb]
	\centering
	\begin{subfigure}[b]{0.49\textwidth}
		\centering
		\includegraphics[width=\linewidth]{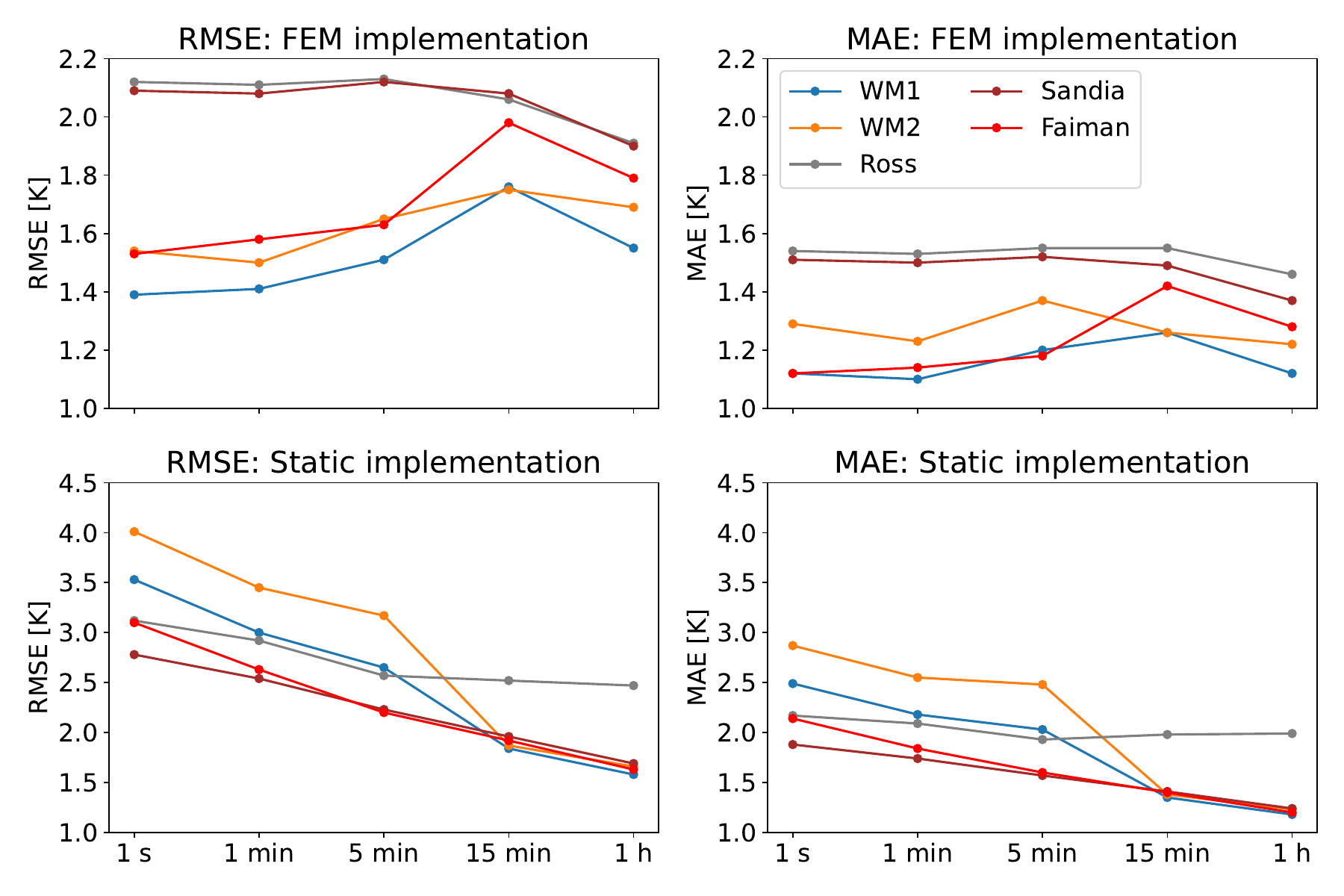}
		\caption{KUL, rooftop array.}
	\end{subfigure}
	\hfill
	\begin{subfigure}[b]{0.49\textwidth}
		\centering
		\includegraphics[width=\textwidth]{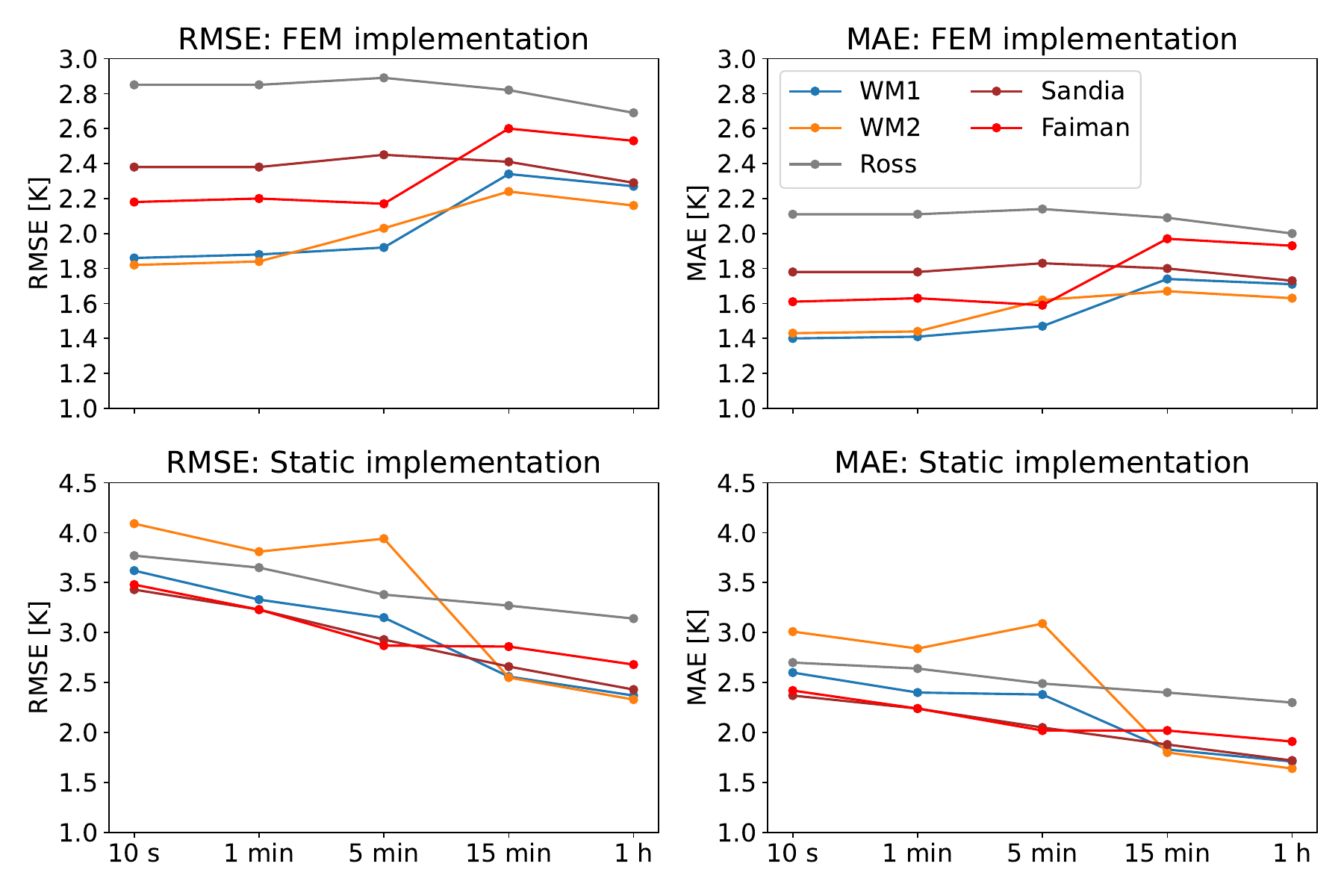} 
		\caption{NIST ground-mount RTD 4.}
	\end{subfigure}
	\caption{Comparison of the FEM implementation versus the static formulation fitted to the data for the five evaluated models, using data from KUL (rooftop array) and NIST (ground-mount, RTD 4). No value for $\tau$ was determined at \SI{1}{h}; hence no EWM impact on this data.}
	\label{Fig:EWM_vs_std_formulation_comparison_1s_to_1h}
	\hfill
\end{figure*}

\Cref{Fig:EWM_vs_std_formulation_comparison_1s_to_1h} shows that the RMSE and MAE of the static models worsen when the time resolution becomes shorter, with pronounced impacts at short timescales, as the irradiance and wind speed signals are then translated in the static model without time delay, which can result in impossible temperature changes or unnecessary noise, especially for sub-minute timescales. By contrast, the FEM approach typically improves the RMSE and MAE values of these models as time steps become shorter. In the non-FEM framework, WM1 and WM2 see the most dramatic worsening of RMSE at shorter timescales, which most likely is a consequence of the more aggressive coefficients used, compared to the other models. The \SI{\leq 1}{min} averaged data for both sites are given in \Cref{Tab:KUL_NIST_leq_1T_RMSE_MAE_vals}, which show that both sites see an average RMSE and MAE improvement from the standard to FEM methodology of \SI{-1.3}{K} (\SI{-40}{\%}) and \SI{-0.9}{K} (\SI{-37}{\%}) respectively.

\newcolumntype{X}{>{\centering\arraybackslash}p{0.05\linewidth}}
\newcolumntype{Y}{>{\centering\arraybackslash}p{0.16\linewidth}}
\newcolumntype{Z}{>{\centering\arraybackslash}p{0.075\linewidth}}

\begin{table*}[h!]
	\centering 
	\caption{RMSE and MAE value comparison for \SI{\leq 1}{min} averaged data for the KUL rooftop and NIST ground-mount RTD 4 systems as seen in \Cref{Fig:EWM_vs_std_formulation_comparison_1s_to_1h}. std = standard methodology, FEM = Filter-EWM-MBE correction method.}
	\label{Tab:KUL_NIST_leq_1T_RMSE_MAE_vals}

	\begin {tabular}{XYZZZcZZ}%
	\toprule \multicolumn {1}{c}{Org}&\multicolumn {1}{c}{KPI}&\multicolumn {1}{c}{WM1}&\multicolumn {1}{c}{WM2}&\multicolumn {1}{c}{Ross}&\multicolumn {1}{c}{Sandia}&\multicolumn {1}{c}{Faiman}&\multicolumn {1}{c}{Average}\\\hline %
	KUL&RMSE std [K]&3.26&3.73&3.02&2.66&2.86&3.11\\%
	KUL&RMSE FEM [K]&1.4&1.52&2.12&2.08&1.56&1.74\\%
	KUL&$\Delta $RMSE [K]&-1.86&-2.21&-0.9&-0.58&-1.31&-1.37\\%
	KUL&$\Delta $RMSE [\%]&-57.1&-59.2&-30.0&-21.6&-45.7&-44.2\\\hline %
	NIST&RMSE std [K]&3.48&3.95&3.71&3.33&3.36&3.56\\%
	NIST&RMSE FEM [K]&1.87&1.83&2.85&2.38&2.19&2.22\\%
	NIST&$\Delta $RMSE [K]&-1.6&-2.12&-0.86&-0.95&-1.16&-1.34\\%
	NIST&$\Delta $RMSE [\%]&-46.2&-53.7&-23.2&-28.5&-34.7&-37.6\\\hline %
	KUL&MAE std [K]&2.34&2.71&2.13&1.81&1.99&2.19\\%
	KUL&MAE FEM [K]&1.11&1.26&1.54&1.5&1.13&1.31\\%
	KUL&$\Delta $MAE [K]&-1.22&-1.45&-0.59&-0.31&-0.86&-0.89\\%
	KUL&$\Delta $MAE [\%]&-52.5&-53.5&-27.9&-16.9&-43.2&-40.4\\\hline %
	NIST&MAE std [K]&2.5&2.92&2.67&2.31&2.33&2.55\\%
	NIST&MAE FEM [K]&1.4&1.44&2.11&1.78&1.62&1.67\\%
	NIST&$\Delta $MAE [K]&-1.1&-1.49&-0.56&-0.53&-0.71&-0.88\\%
	NIST&$\Delta $MAE [\%]&-43.8&-50.9&-21.0&-22.8&-30.5&-34.4\\\bottomrule %
	\end {tabular}%

\end{table*}

\subsection{Impact of filtering, EWM and MBE methods}\label{Subsec:Results_FEM_impact}

\begin{figure*}[htb]
	\centering
	\includegraphics[width=\linewidth]{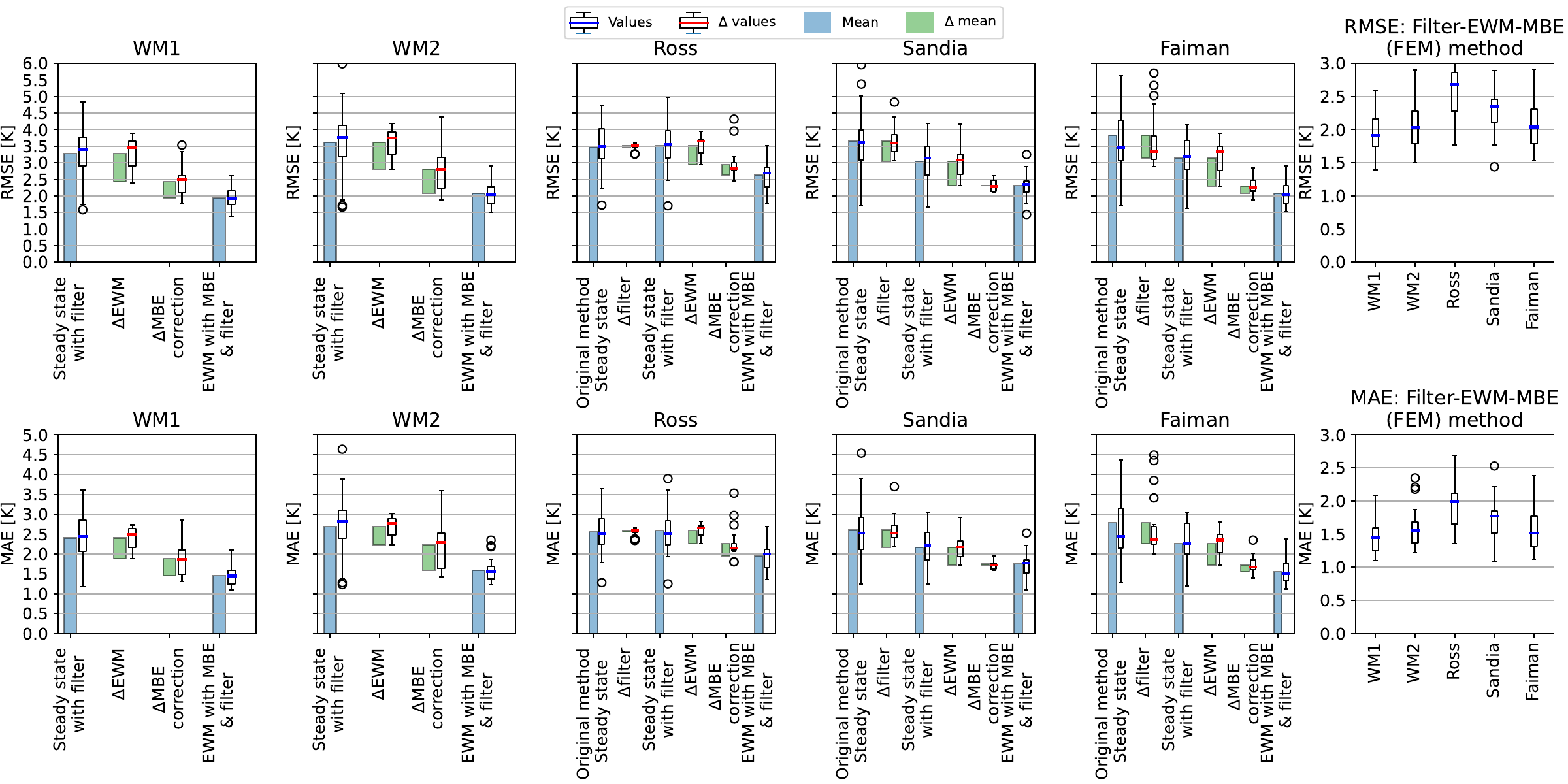}
	\caption{Waterfall and boxplot charts for the five thermal models, evaluated on the datasets; top row is RMSE, bottom row is MAE. The boxplots show the spread around the mean values used in the waterfall, while the waterfall shows the average trend and impact per category. Boxplot outliers are marked when data are beyond the whiskers, which extend 1.5 times the inter-quartile range, shown by the box. The median value is marked for each boxplot; results are shown in blue, and intermediate steps, such as the impact of the EWM approach, are red.}
	\label{Fig:RMSE_MAE_Waterfall_boxplot_all}
\end{figure*}

\Cref{Fig:RMSE_MAE_Waterfall_boxplot_all} shows that all models benefit from filtering, EWM, and MBE correction. The extent to which each site and model benefits varies, with an average improvement for all datasets for RMSE of \SIrange{-0.7}{-1.5}{K} (\SIrange{-20}{-40}{\%}) and MAE of \SIrange{-0.4}{-1.1}{K} (\SIrange{-20}{-40}{\%}). All model results for the FEM versus the standard approach can be found in the appendix, in \Cref{Tab:all RMSE values for FEM vs std,Tab:all MAE values for FEM vs std,Tab:all MBE values for FEM vs std}.
In \Cref{Fig:RMSE_MAE_Waterfall_boxplot_all}, the \textit{mean} RMSE benefit per category hides significant variations that are observed per model. Given that Ross's model works with no knowledge of wind speeds, it benefits less from the FEM approach than the other models which do incorporate wind speeds. Overall, the overwhelming majority of the 15 sites and 24 datasets see an improvement in the final Filtered-EWM-MBE error metrics, for all five models tested. The average standard deviation $\sigma$ in the FEM RMSE and MAE values versus the standard approach is halved, which is of importance, as this reduces the uncertainty of the thermal model results, thus giving increased confidence for financing of PV plants.  

\newcolumntype{K}{>{\centering\arraybackslash}p{0.05\linewidth}}
\newcolumntype{L}{>{\centering\arraybackslash}p{0.20\linewidth}}
\newcolumntype{M}{>{\centering\arraybackslash}p{0.065\linewidth}}

\begin{table*}[htb]
	\centering 
	\caption{Average standard and FEM methodology RMSE and MAE values for all models and datasets as shown in \Cref{Fig:RMSE_MAE_Waterfall_boxplot_all}, with additional MBE data.}
	\label{Tab:all_models_all_sites_avg_RMSE_MAE}
	\begin {tabular}{cKKKcMM}%
	\toprule Model&WM1&WM2&Ross&Sandia&Faiman&Average\\\hline %
	RMSE$_{std}$ [K]&3.29&3.62&3.52&3.04&3.15&3.32\\%
	RMSE$_{FEM}$ [K]&1.93&2.08&2.62&2.32&2.08&2.21\\%
	$\Delta $RMSE [K]&-1.36&-1.54&-0.9&-0.72&-1.07&-1.12\\%
	$\Delta $RMSE [\%]&-41.3&-42.5&-25.6&-23.7&-34.0&-33.4\\%
	$\sigma _{RMSE-std}$ [K]&0.84&1.05&0.78&0.7&0.74&0.82\\%
	$\sigma _{RMSE-FEM}$ [K]&0.34&0.38&0.45&0.36&0.39&0.38\\\hline %
	MAE$_{std}$ [K]&2.4&2.69&2.59&2.17&2.26&2.42\\%
	MAE$_{FEM}$ [K]&1.46&1.59&1.95&1.75&1.56&1.66\\%
	$\Delta $MAE [K]&-0.94&-1.1&-0.64&-0.42&-0.7&-0.76\\%
	$\Delta $MAE [\%]&-39.2&-40.9&-24.7&-19.4&-31.0&-31.0\\%
	$\sigma _{MAE-std}$ [K]&0.63&0.82&0.6&0.52&0.54&0.62\\%
	$\sigma _{MAE-FEM}$ [K]&0.27&0.31&0.34&0.31&0.34&0.31\\\hline %
	MBE$_{std}$ [K]&-0.84&-1.06&-0.57&0.17&-0.23&-0.51\\%
	MBE$_{FEM}$ [K]&0.11&0.11&0.13&0.13&0.11&0.12\\%
	$\Delta | \text {MBE}|$ [K]&-0.73&-0.95&-0.44&-0.04&-0.12&-0.46\\\bottomrule %
	\end {tabular}%
	
\end{table*}

\subsection{Impact on energy error estimates and the importance of MAE}

In the literature, the translation of the model error into power error is taken as $RMSE \cdot \gamma$, with $\gamma$ the module's coefficient of power \cite{Faiman_thermal_model_PiP_2008}. If this method is applied to the module energy, an estimation error can be made. 

With \Cref{Eq:Absolute energy error year}, it is possible to estimate the absolute and relative energy estimation error, due to the temperature model error, for a given time resolution. As an \textit{approximation}, \Cref{Eq:Absolute energy error year_approx,Eq:PR estimation error year approx} can serve to quantify the estimation error. 
\begin{align}
	\Delta E(|\Delta T_{o,e}|) &= \sum_{0}^{N} G_i \cdot \big|T_{o,model} - T_{o,meas}\big|_i \cdot \gamma \nonumber \\ & \bigg[\frac{kWh}{kWp\cdot y}\bigg]	\label{Eq:Absolute energy error year} \\
\Delta E(|\Delta T_{o,e}|) &\approx H_{y}\cdot MAE \cdot \gamma \quad \bigg[\frac{kWh}{kWp\cdot y}\bigg] \label{Eq:Absolute energy error year_approx} \\
\Delta PR(\Delta T_{o,e})	&\approx MAE \cdot \gamma \quad \bigg[ \frac{\%}{y} \bigg] \label{Eq:PR estimation error year approx}
\end{align}

The directional error (energy \textit{over-}, respectively \textit{under-} \textit{estimation} due to temperature \textit{under-}, respectively \textit{over-} \textit{prediction}) can be found by using \Cref{Eq:Energy error under-estimate,Eq:Energy error over-estimate}. 
\begin{align}
	\Delta E(\Delta T_{o,e<0}) &= \sum_{0}^{N} G_i \cdot \big(T_{o,model} - T_{o,meas}\big)_i^{<0} \cdot \gamma
	\label{Eq:Energy error over-estimate} \\
	\Delta E(\Delta T_{o,e>0}) &= \sum_{0}^{N} G_i \cdot \big(T_{o,model} - T_{o,meas}\big)_i^{>0} \cdot \gamma \label{Eq:Energy error under-estimate} 
\end{align}

\begin{figure}[htb]
	\centering
	\includegraphics[width=1\linewidth]{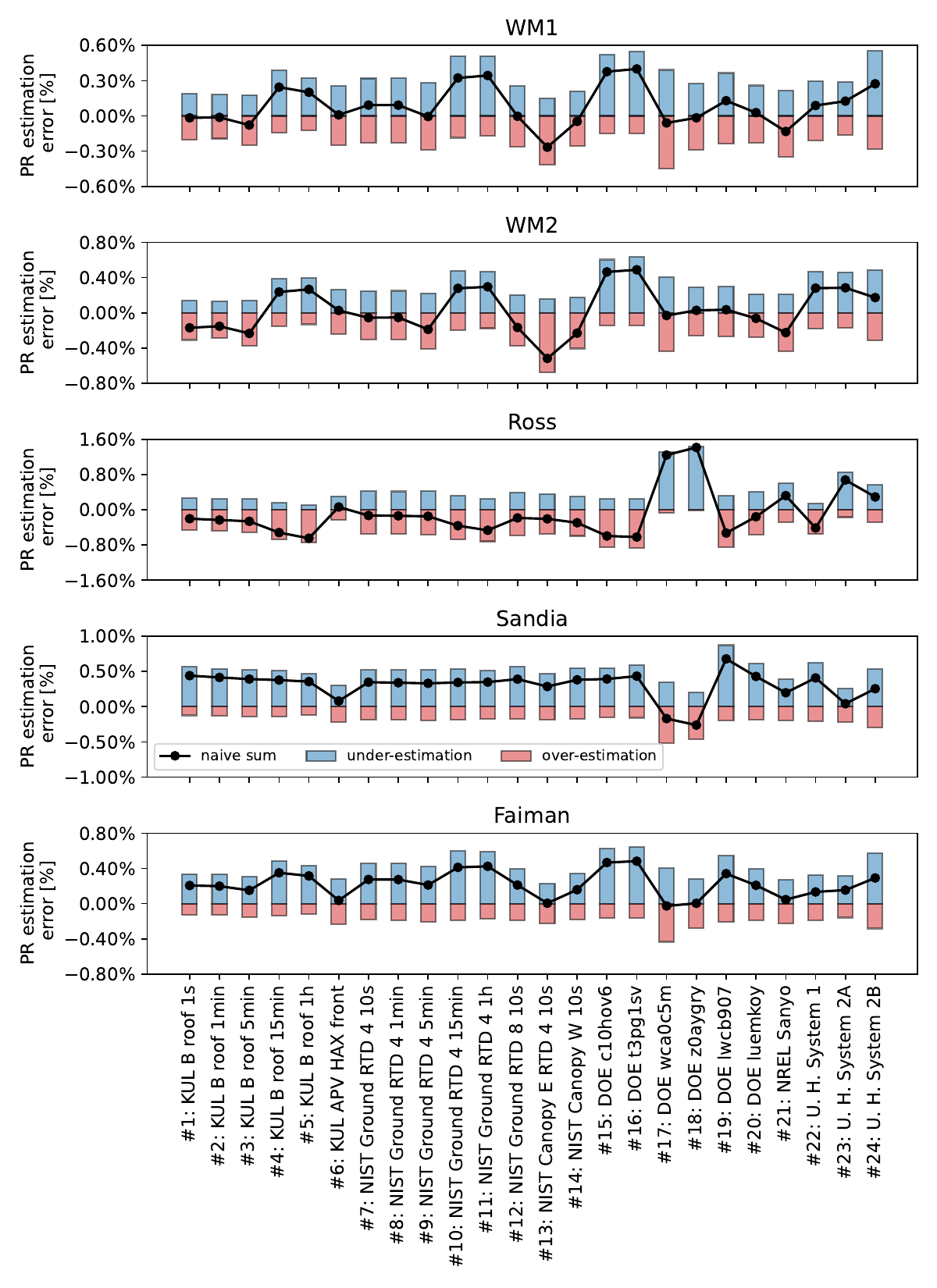}
	\caption{PR estimation error, using $\gamma =$ \SI{-0.35}{\percent/K}. A model under- (over-) estimation occurs when the modelled temperature is lower (higher) than the measured value. The spread between under- and over-estimation error is characterised by the MAE.}
	\label{Fig:PR_estimation_error}
\end{figure}

For example, for the KUL Ghent rooftop array at \SI{1}{s} resolution, taking a thermal coefficient of power $\gamma$ at \SI{-0.35}{\percent/K}, a \textit{total} energy error of \SI{4.09}{kWh/kWp} (\SI{0.39}{\%}-points on PR) is found, with \SI{1.96}{kWh/kWp} over-estimate and \SI{-2.13}{kWh/kWp} under-estimate, resulting in an naive error of \SI{-0.17}{kWh/kWp}, as most of the temperature model errors cancel out over the year. A near-zero naive error does not apply to all sites and models in their FEM form, as evidenced in \Cref{Fig:PR_estimation_error}. For the sites and datasets where both WM1 and WM2 under-estimate the module temperature, this suggests that the coefficient $k$ is too low. 
The spread between under- and over-estimation error is characterised by the MAE, highlighting its importance as a model error KPI.  

Note also that software that use longer time steps for the energy yield forecasts, e.g. \SI{1}{h} time steps for one-year calculations, may mis-estimate the temperature impact on the yield, due to the model error (which itself is impacted by the chosen time resolution). As seen in \Cref{Fig:EWM_vs_std_formulation_comparison_1s_to_1h}, the RMSE and MAE values in the non-FEM formulation are approximately \SIrange{0.8}{1.5}{K} higher than at \SI{1}{h} for \SIrange{1}{5}{min} data. Consequently, the temperature impact on the energy yield is also mis-identified. This is particularly important for the (initial) operational phase of the PV power plant and the stakeholders involved, as the MAE on the thermal model can thus result in a PR error of \SIrange{0.5}{1.5}{pp}.

\subsection{Contextualisation: model error versus module sensors}

So far, the discussion has focused on the resulting error metrics on their own. However, it is useful to contextualise these results against measured data. One such thought experiment that can be done is:
\begin{quote} ``How large is the error, if the data from the (perfect) sensor is delayed by \SIrange{1}{10}{min}?''. \end{quote}
 
For this, the sensor data is compared against its time-delayed value. In the case of the KUL Ghent roof B site, the module used has multiple RTD sensors laminated against the cells, with two backsheet sensors next to these. Thus, a backsheet-to-cell (BS-C) correction can be done as per \cite{Sandia_King_PV_array_perf_model_2004} and \Cref{Eq:CBS correction}, with the added modification that the irradiance signal again has the EWM methodology applied to it, and $ k_{BS-C}$ the R-value for the backsheet-to-cell difference.

\begin{equation}
T_{cell,BS-C} = T_{backsheet} + G_{EWM}\cdot k_{BS-C} \: [K]
\label{Eq:CBS correction}
\end{equation}

\begin{figure}[htb]
	\includegraphics[width=\linewidth]{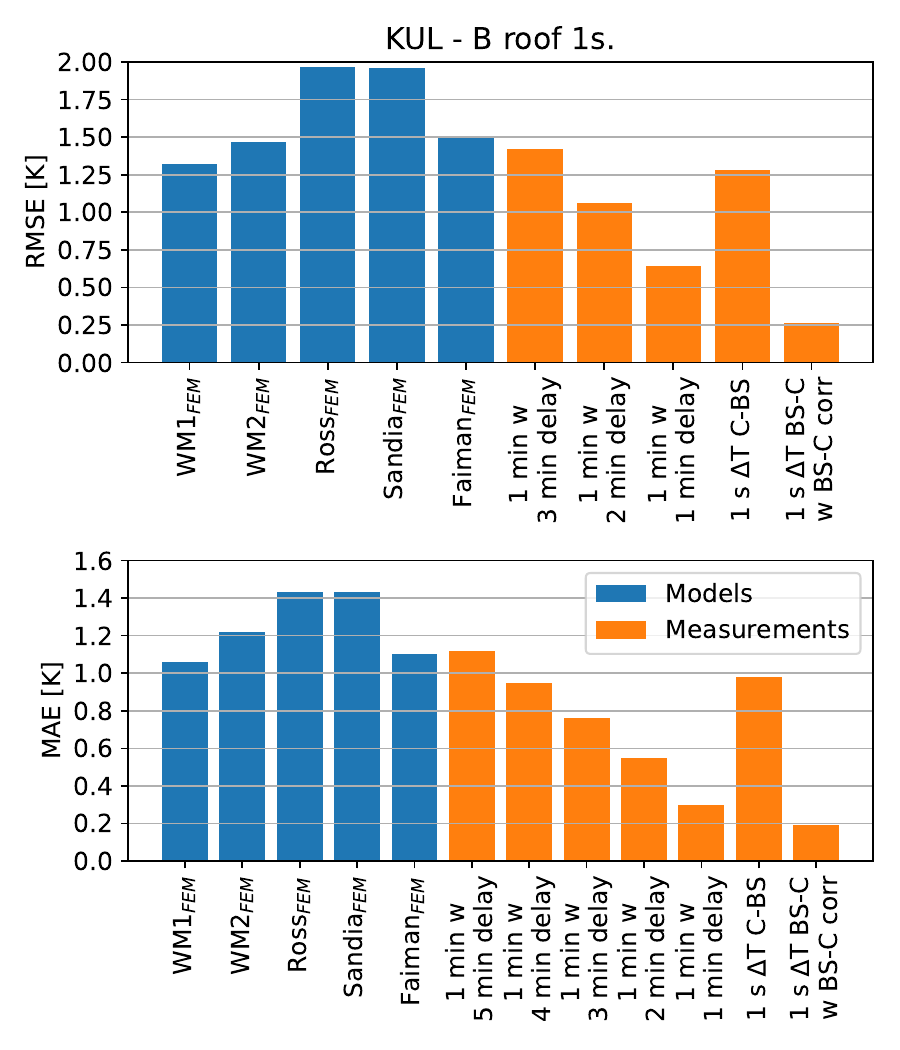}
\caption{KUL Ghent: RMSE and MAE values of models, versus time-delayed measured data, including backsheet data with backsheet-to-cell correction.}
\label{Fig:contextualising model errors vs sensor measurements}
\end{figure}

\Cref{Fig:contextualising model errors vs sensor measurements} shows that, while the RMSE and MAE values for the models which incorporate wind speeds (WM1, WM2, King and Faiman) have improved significantly through the use of the FEM methodology, the sensor on the backsheet provides an even better result. Applying the backsheet-to-cell correction results in an MAE and RMSE of approximately \SI{0.2}{K} and \SI{0.25}{K} respectively, which is well below the uncertainty of the RTD sensors used (approximately \SI{0.4}{K} at k=2). In practical terms, this means that (well-installed and maintained) module sensors remain the preferred option against using a modelled value only, yet the thermal model can be used, among others to verify the quality of the sensor over time, as well as to estimate or forecast future temperature values. Contextualised versus time, the best models (at \SI{1}{s} resolution) give approximately the same result as a ``true'' \SI{1}{min} measurement point, delayed by \SI{3}{min} (RMSE) and \SI{5}{min} (MAE).

\subsection{Discussion}

From \Cref{Fig:PR_estimation_error,Fig:contextualising model errors vs sensor measurements} we are confronted with a sobering result: despite the significant improvement to all of the models for nearly all of the datasets tested in this work by using the FEM methodology, much work remains to be done for thermal models to approach the accuracy of measured data. Potential avenues to explore are to include the relative humidity in the thermal model(s), as done by \cite{Veldhuis_Peters_EWM_model_2015,Peters_Thermal_Floating_PV_time_const_2022}. Investigating the impact of wind direction and/or local turbulence effects may also further reduce model errors, albeit likely at high model complexity or computational cost. Additionally, correcting for precipitation events (e.g. by setting $T_o = 0$ during heavy rain) can reduce temperature model errors (although this changes little to the fundamental model behaviour and its KPIs during times without rain). 

The conceptual model and theoretical calculations versus measured data summarised in \Cref{Tab:R_C_tau_theoretical,Tab:r_c_tau_results} indicate that while wind access to the module surface(s) can be very important, the fact that heat removal paths occur via the front and back of the module which are in parallel to the heat source, sees a more limited effect than otherwise expected. For example, if wind hits the front glass and reduces the effective wind film layer thickness, the back is affected very little, leaving a high R-value in place, and vice versa for wind from the back. Nevertheless, this also suggests that local obstructions will also result in wind direction effects, e.g. for a row of trees on one side of an array, up to wind barrier effects from rows of modules for utility-scale PV farms. As such, the wind speed signal on its own does not suffice when attempting to reach RMSE and MAE values below \SI{1}{K}. Quantifying the magnitude of such wind direction effects still needs to be done, and verifying such data using finite element analysis and wind tunnel experiments are interesting avenues for further work.

An under-appreciated nuance in the PV field by different stakeholders (financiers, asset owners/investors, and engineering-procurement-construction (EPC) companies) in using (thermal) models ex-ante (i.e. forecast, with zero measurement data, prior to system construction) for ex-post evaluation (i.e. comparison of measured data of the built PV system against the thermal model). As seen in \Cref{Subsec:FEM_vs_std_timescales,Subsec:Results_FEM_impact}, thermal models can vary significantly for different timescales, and local wind access to modules is estimated, based on best available knowledge. Historically, most ``bankable'' PV modelling software packages output hourly data, which then form part of the contractual model for a PV system. Performance ratio calculations, yield comparisons, and model corrections (e.g. for different weather conditions versus the contractual model) are thus placed at that time resolution, whereas the true performance of the system has in reality evolved in much shorter time steps, subject to varying weather conditions: wind gusts and lulls, irradiance peaks and troughs, ambient temperature, humidity, and precipitation. As discussed previously, the \textit{sequence} of weather events plays a material role in the eventual temperature of the PV module, whereas such results are hidden (or lost) when an hourly resolution is employed \cite{Ransome_hourly_data_insufficient_PVSEC_2005,Riley_quantifying_time_effects_averaging_sampling_PVSC_2009}.

If the MAE and RMSE of the thermal model vary (strongly) with time resolution, this also impacts the attribution or estimation of effects on a PV system's performance, which can be at least \SIrange{0.5}{1.5}{pp} on the PR. The asset owner and EPC (and/or operations and maintenance provider) may thus be presented with a quandary: a PV system may be under-performing compared to the guaranteed yield or PR, yet the contractual model with measured weather data (at \SI{1}{h} resolution) appears to meet expectations.

At such a moment, the need arises for clear attribution of effects on system performance, for which higher-resolution data must be used, where the FEM approach can aid to estimate thermal effects. A further point which may be subject to discussion is whether thermal model coefficients used for the contractual model are fixed (which may still be the case for contractual discussions), while an updated energy estimation model with coefficients from measured data can be used for the asset owner financial forecasts.

An additional key point to consider is that the model coefficients are determined in this work for data of PV systems that see limited to zero power-constrained conditions, such as inverter clipping or mandated curtailment. The linearisation approach used in this work for coefficient determination will not work as well for power-constrained conditions, which depend on the system configuration and local grid conditions, as well as the local weather. Given the trends for increased DC-AC ratios of PV systems, as well as mandated curtailment due to increasing Renewable Power Fractions \cite{Herteleer_Visions_future_curtailment_flexibility_PVSEC_2018,Shaughnessy_PV_Global_Trends_2020}, this is a challenge for thermal models that is yet to be addressed.

\section{Conclusions}

This paper has presented a framework for thermal models, based on an RC-equivalent conceptual thermal model of a PV module. Using this conceptual model, it is possible to determine the equivalent thermal resistance and capacitance of a PV module, as well as the thermal time constant. 

Informed by the RC-equivalent thermal model, improved filtering approaches to determine coefficients were shown (particularly for low and high wind speeds), which form the first step in the filter-EWM-MBE correction (FEM) methodology. This then also facilitates the determination of the module thermal time constant $\tau$, which is shown to vary in function of the wind speed. The exponential weighted mean (EWM) calculation of irradiance and wind speed signals can then be calculated using the time step $\Delta t$ and $\tau$. The EWM step makes a static model dynamic. By then using the Mean Bias Error of the training dataset as a fixed radiation loss value, the FEM methodology is complete. 

With $r_{eq,max} = k = e^a = \frac{1}{U_0}$, two thermal models (WM1 and WM2) are proposed, where WM1 is a modified form of the Sandia model, with the coefficients for WM1 determined in two regression steps, compared to the single regression used for the Sandia model. The four models which incorporate the wind speed (WM1, WM2, Sandia, and Faiman), reduce to Ross' model when the wind speed is equal to zero.

Despite the significant improvement to all of the models for nearly all of the datasets tested in this work by using the FEM methodology (with an average improvement for all models and datasets for RMSE of \SIrange{-0.7}{-1.5}{K} (\SIrange{-20}{-40}{\%}) and MAE of \SIrange{-0.4}{-1.1}{K} (\SIrange{-20}{-40}{\%}), much work remains to be done for thermal models to approach the accuracy of measured data. As such, having well-installed and maintained module temperature sensors should be the preferred option, yet the FEM methodology can be used for pre- and post-installation temperature estimations, and used among others to monitor sensor quality. The EWM step is particularly useful for pre-installation and design modelling. The average standard deviation $\sigma$ in the FEM RMSE and MAE values is halved versus the standard approach, which is of importance, as this reduces the uncertainty of the thermal model results, thus giving increased confidence for financing of PV plants which apply the FEM methodology.

\section*{Acknowledgements}
This work was supported by the SOLARISE project. The SOLARISE project received funding from the Interreg 2 Seas programme 2014-2020 co-funded by the European Regional Development Fund under subsidy contract N° 2S04-004, additionally co-funded by Province of East Flanders. 


\section*{Data availability}
Datasets related to this article can be found at
\begin{itemize}
	\item \url{https://doi.org/10.48804/RVTSD4}, hosted by KU Leuven RDR \cite{KUL_Rooftop_array_1s_weather_dataset_2015_2016}
	\item \url{https://doi.org/10.18434/M3S67G}, hosted by NIST \cite{NIST_PV_dataset};
	\item \url{https://doi.org/10.17605/OSF.IO/VTR2S}, hosted by OSF \cite{T13_data};
	\item \url{https://doi.org/10.5281/zenodo.3958820}, hosted by Zenodo \cite{Barry_dyn_thermal_model_data_2020};
	
\end{itemize}

\bibliography{PV_thermal_models}

\section*{Appendix}\label{sec:Appendix}

\newcolumntype{P}{>{\centering\arraybackslash}p{0.12\linewidth}}
\newcolumntype{Q}{>{\centering\arraybackslash}p{0.055\linewidth}}

\begin{table*}[tb]
	\centering 
	\caption{RMSE model results for all sites and time resolutions, FEM approach versus standard.}
	\label{Tab:all RMSE values for FEM vs std}
	\begin {tabular}{Qc|ccccc|ccccc}%
	\toprule Org & Site & \multicolumn {5}{c}{FEM}&\multicolumn {5}{c}{standard}\\%
	&&WM1&WM2&Ross&Sand&Faim&WM1&WM2&Ross&Sand&Faim\\\hline %
	KUL&B roof 1s&1.39&1.54&2.12&2.09&1.53&3.53&4.01&3.12&2.78&3.1\\%
	KUL&B roof 1min&1.41&1.5&2.11&2.08&1.58&3.0&3.45&2.92&2.54&2.63\\%
	KUL&B roof 5min&1.51&1.65&2.13&2.12&1.63&2.65&3.17&2.57&2.23&2.2\\%
	KUL&B roof 15min&1.76&1.75&2.06&2.08&1.98&1.84&1.87&2.52&1.96&1.92\\%
	KUL&B roof 1h&1.55&1.69&1.91&1.9&1.79&1.58&1.66&2.47&1.69&1.63\\%
	KUL&APV HAX front&1.77&1.79&1.77&1.77&1.78&1.74&1.71&1.7&1.67&1.69\\%
	NIST&Ground RTD 4 10s&1.86&1.82&2.85&2.38&2.18&3.62&4.09&3.77&3.43&3.48\\%
	NIST&Ground RTD 4 1min&1.88&1.84&2.85&2.38&2.2&3.33&3.81&3.65&3.23&3.23\\%
	NIST&Ground RTD 4 5min&1.92&2.03&2.89&2.45&2.17&3.15&3.94&3.38&2.93&2.87\\%
	NIST&Ground RTD 4 15min&2.34&2.24&2.82&2.41&2.6&2.56&2.55&3.27&2.66&2.86\\%
	NIST&Ground RTD 4 1h&2.27&2.16&2.69&2.29&2.53&2.37&2.33&3.14&2.43&2.68\\%
	NIST&Ground RTD 8 10s&1.91&2.0&2.94&2.58&2.16&4.33&4.99&4.04&3.7&3.93\\%
	NIST&Canopy E RTD 4 10s&1.98&2.68&2.68&2.15&1.72&4.85&5.99&3.69&3.43&4.01\\%
	NIST&Canopy W 10s&1.7&1.98&2.59&2.34&1.84&4.33&5.09&3.94&3.72&3.93\\%
	DOE&c10hov6&2.44&2.7&3.52&2.51&2.79&3.43&3.72&4.45&3.41&3.67\\%
	DOE&t3pg1sv&2.6&2.9&3.51&2.75&2.91&3.5&3.81&4.54&3.53&3.69\\%
	DOE&wca0c5m&2.36&2.34&2.79&2.48&2.34&4.07&4.12&4.81&4.18&4.1\\%
	DOE&z0aygry&2.05&2.04&2.57&2.36&2.05&4.14&4.23&4.98&4.17&4.15\\%
	DOE&lwcb907&2.13&2.05&2.98&3.25&2.47&3.97&4.19&4.21&3.67&3.68\\%
	DOE&luemkoy&1.78&1.78&2.74&2.45&1.97&3.36&3.56&3.53&3.1&3.14\\%
	NREL&Sanyo&2.05&2.26&3.01&2.29&1.9&3.72&4.12&3.36&2.94&3.14\\%
	U. H.&System 1&2.01&2.57&2.37&2.89&2.03&3.57&3.54&3.59&3.48&3.56\\%
	U. H.&System 2A&1.48&2.47&2.63&1.44&1.55&3.02&3.74&3.63&2.98&3.1\\%
	U. H.&System 2B&2.26&2.1&2.33&2.25&2.3&3.24&3.18&3.2&3.18&3.21\\%
	\hline Avg&all&1.93&2.08&2.62&2.32&2.08&3.29&3.62&3.52&3.04&3.15\\\bottomrule %
	\end {tabular}%
	
\end{table*}

\begin{table*}[tb]
	\centering 
	\caption{MAE model results for all sites and time resolutions, FEM approach versus standard.}
	\label{Tab:all MAE values for FEM vs std}
	\begin {tabular}{Qc|ccccc|ccccc}%
	\toprule Org & Site & \multicolumn {5}{c}{FEM}&\multicolumn {5}{c}{standard}\\%
	&&WM1&WM2&Ross&Sand&Faim&WM1&WM2&Ross&Sand&Faim\\\hline %
	KUL&B roof 1s&1.12&1.29&1.54&1.51&1.12&2.49&2.87&2.17&1.88&2.14\\%
	KUL&B roof 1min&1.1&1.23&1.53&1.5&1.14&2.18&2.55&2.09&1.74&1.84\\%
	KUL&B roof 5min&1.2&1.37&1.55&1.52&1.18&2.03&2.48&1.93&1.57&1.6\\%
	KUL&B roof 15min&1.26&1.26&1.55&1.49&1.42&1.35&1.38&1.98&1.41&1.4\\%
	KUL&B roof 1h&1.12&1.22&1.46&1.37&1.28&1.18&1.23&1.99&1.24&1.2\\%
	KUL&APV HAX front&1.35&1.37&1.36&1.36&1.36&1.31&1.28&1.25&1.24&1.27\\%
	NIST&Ground RTD 4 10s&1.4&1.43&2.11&1.78&1.61&2.6&3.01&2.7&2.37&2.42\\%
	NIST&Ground RTD 4 1min&1.41&1.44&2.11&1.78&1.63&2.4&2.84&2.64&2.24&2.24\\%
	NIST&Ground RTD 4 5min&1.47&1.62&2.14&1.83&1.59&2.38&3.09&2.49&2.05&2.02\\%
	NIST&Ground RTD 4 15min&1.74&1.67&2.09&1.8&1.97&1.83&1.8&2.4&1.88&2.02\\%
	NIST&Ground RTD 4 1h&1.71&1.63&2.0&1.73&1.93&1.71&1.64&2.3&1.72&1.91\\%
	NIST&Ground RTD 8 10s&1.45&1.59&2.19&1.91&1.56&3.15&3.73&2.87&2.54&2.73\\%
	NIST&Canopy E RTD 4 10s&1.55&2.18&2.03&1.64&1.29&3.61&4.64&2.63&2.38&2.83\\%
	NIST&Canopy W 10s&1.32&1.58&1.99&1.8&1.39&3.2&3.89&2.83&2.6&2.78\\%
	DOE&c10hov6&1.99&2.21&2.67&2.04&2.32&2.52&2.8&3.33&2.54&2.78\\%
	DOE&t3pg1sv&2.09&2.35&2.69&2.22&2.38&2.58&2.87&3.44&2.65&2.8\\%
	DOE&wca0c5m&1.84&1.82&2.2&1.94&1.82&2.98&3.05&3.62&3.05&3.04\\%
	DOE&z0aygry&1.49&1.47&1.99&1.76&1.49&3.03&3.15&3.9&3.01&3.06\\%
	DOE&lwcb907&1.45&1.41&2.11&2.53&1.77&3.06&3.25&3.3&2.81&2.82\\%
	DOE&luemkoy&1.21&1.24&1.95&1.83&1.36&2.37&2.52&2.53&2.18&2.19\\%
	NREL&Sanyo&1.48&1.67&2.17&1.61&1.33&2.81&3.17&2.41&2.09&2.28\\%
	U. H.&System 1&1.54&1.87&1.81&2.21&1.54&2.53&2.44&2.5&2.58&2.49\\%
	U. H.&System 2A&1.13&1.72&1.95&1.09&1.2&2.07&2.73&2.64&2.0&2.15\\%
	U. H.&System 2B&1.73&1.52&1.69&1.68&1.77&2.31&2.25&2.26&2.25&2.29\\%
	\hline Avg&all&1.46&1.59&1.95&1.75&1.56&2.4&2.69&2.59&2.17&2.26\\\bottomrule %
	\end {tabular}%
	
\end{table*}

\begin{table*}[tb]
	\centering 
	\caption{MBE model results for all sites and time resolutions, FEM approach versus standard.}
	\label{Tab:all MBE values for FEM vs std}
	\begin {tabular}{Qc|ccccc|ccccc}%
	\toprule Org & Site & \multicolumn {5}{c}{FEM}&\multicolumn {5}{c}{standard}\\%
	&&WM1&WM2&Ross&Sand&Faim&WM1&WM2&Ross&Sand&Faim\\\hline %
	KUL&B roof 1s&-0.01&0.02&0.25&0.09&-0.02&-1.71&-2.21&-1.08&0.01&-0.96\\%
	KUL&B roof 1min&-0.01&0.01&0.24&0.08&-0.02&-1.53&-2.02&-1.08&0.04&-0.78\\%
	KUL&B roof 5min&-0.02&0.0&0.24&0.09&-0.02&-1.64&-2.21&-1.08&0.05&-0.81\\%
	KUL&B roof 15min&0.0&-0.03&0.24&0.08&0.01&-0.46&-0.44&-1.51&0.0&-0.08\\%
	KUL&B roof 1h&0.01&-0.06&0.24&0.06&0.01&-0.65&-0.3&-1.75&-0.07&-0.21\\%
	KUL&APV HAX front&-0.09&-0.09&-0.06&-0.07&-0.09&-0.3&-0.23&-0.0&0.04&-0.2\\%
	NIST&Ground RTD 4 10s&0.17&0.16&0.08&0.15&0.17&-1.27&-2.02&-0.7&0.16&-0.4\\%
	NIST&Ground RTD 4 1min&0.17&0.17&0.08&0.15&0.17&-1.21&-1.98&-0.7&0.17&-0.31\\%
	NIST&Ground RTD 4 5min&0.17&0.17&0.09&0.15&0.17&-1.67&-2.64&-0.7&0.19&-0.58\\%
	NIST&Ground RTD 4 15min&0.16&0.17&0.09&0.15&0.15&0.05&-0.28&-1.14&0.16&0.51\\%
	NIST&Ground RTD 4 1h&0.15&0.17&0.08&0.15&0.14&0.06&-0.29&-1.37&0.06&0.49\\%
	NIST&Ground RTD 8 10s&0.23&0.22&0.14&0.2&0.23&-2.1&-2.95&-0.91&0.12&-1.09\\%
	NIST&Canopy E RTD 4 10s&0.02&-0.01&-0.11&-0.0&0.02&-2.92&-4.13&-0.74&0.18&-1.59\\%
	NIST&Canopy W 10s&0.07&0.06&-0.03&0.07&0.08&-2.3&-3.22&-1.12&0.06&-1.31\\%
	DOE&c10hov6&0.11&0.18&-0.18&0.11&0.09&0.35&0.96&-1.87&0.43&0.94\\%
	DOE&t3pg1sv&0.15&0.22&-0.15&0.13&0.13&0.21&0.85&-2.0&0.44&0.76\\%
	DOE&wca0c5m&-0.01&-0.01&0.01&-0.0&-0.01&0.96&1.14&2.95&0.27&1.16\\%
	DOE&z0aygry&0.05&0.05&0.07&0.05&0.05&1.88&2.13&3.48&0.29&1.98\\%
	DOE&lwcb907&0.04&0.03&0.35&0.2&0.04&-2.16&-2.51&-2.0&0.1&-1.37\\%
	DOE&luemkoy&-0.05&-0.06&0.25&0.11&-0.04&-1.43&-1.73&-0.9&0.19&-0.76\\%
	NREL&Sanyo&0.05&0.04&0.09&0.05&0.06&-1.86&-2.37&0.41&0.27&-0.88\\%
	U. H.&System 1&0.84&0.83&0.8&0.7&0.81&-1.52&-0.47&-1.5&0.62&-1.45\\%
	U. H.&System 2A&0.15&0.12&0.07&0.21&0.08&0.82&1.85&1.46&0.24&1.05\\%
	U. H.&System 2B&0.33&0.3&0.2&0.28&0.32&0.25&-0.28&0.17&0.13&0.26\\%
	\hline Avg&all&0.11&0.11&0.13&0.13&0.11&-0.84&-1.06&-0.57&0.17&-0.23\\\bottomrule %
	\end {tabular}%
	
\end{table*}

\end{document}